 \documentclass[aps,prfluids,onecolumn,superscriptaddress]{revtex4-2}

\usepackage{graphicx}
\usepackage{amsmath}
\usepackage{amssymb}
\usepackage{xcolor}
\usepackage{tikz}
\usetikzlibrary{shapes}
\usepackage{pgf} 
\usepackage{siunitx}
\usepackage{booktabs} 
\usepackage{multirow} 
\usepackage{bigdelim}
\usepackage{amssymb,pifont}
\usepackage{appendix}
\usepackage{array}

\definecolor{nltw_blue}{rgb}{0,0.6,1}
\definecolor{sw_green}{rgb}{0,0.5647,0}
\definecolor{hj_orange}{rgb}{0.8118,0.3412,0.2745}

\definecolor{color_1}{rgb}{0.9451,0.5569,0.1098}
\definecolor{color_2}{rgb}{0.8824,0.0980,0}
\definecolor{color_3}{rgb}{0,0.0745,0.7490}
\definecolor{color_4}{rgb}{0.7490,0,0.3765}
\definecolor{color_5}{rgb}{0,0.6000,1.0000}
\definecolor{color_6}{rgb}{0.5490,0.7333,0.1490}
\definecolor{color_7}{rgb}{0.0980,0.5020,0}
\definecolor{color_8}{rgb}{0.25,0.25,0.25}

\definecolor{color_LVaries}{rgb}{0.18, 0.52549, 0.9725}
\definecolor{color_DVaries}{rgb}{0.2196, 07843, 0.58823}
\definecolor{color_TauVaries}{rgb}{0.9961, 0.7686, 0.2196}
\definecolor{color_L52}{rgb}{0.2631, 0.149, 0.6588}

\definecolor{color_LVariesContour}{rgb}{0.1098, 0.6667, 0.8745}
\definecolor{color_DVariesContour}{rgb}{0.5882, 07922, 0.2823}
\definecolor{color_TauVariesContour}{rgb}{0.9647, 0.949, 0.1176}
\definecolor{color_L52Contour}{rgb}{0.2823, 0.2863, 0.9294}

\definecolor{color_VelExp}{rgb}{0.039, 0.447, 0.725}
\definecolor{color_Blue}{rgb}{0, 0, 1}
\definecolor{color_Red}{rgb}{1, 0, 0}

\usepackage{subfigure}

\begin{document}

\newcommand{\markerLVariesSimu}{\raisebox{0.5pt}{\tikz{\node[draw,scale=0.5,regular polygon, regular polygon sides=3,fill=color_LVaries,draw=color_LVariesContour](){};}}}
\newcommand{\markerDVariesSimu}{\raisebox{0.5pt}{\tikz{\node[draw,scale=0.5,regular polygon, regular polygon sides=3,fill=color_DVaries,draw=color_DVariesContour, rotate = 180](){};}}}
\newcommand{\markerTauVariesSimu}{\raisebox{0.5pt}{\tikz{\node[draw,scale=0.6,regular polygon, regular polygon sides=4,fill=color_TauVaries,draw=color_TauVariesContour,rotate=45](){};}}}
\newcommand{\markerLFiveTwoSimu}{\raisebox{0.5pt}{\tikz{\node[draw,scale=0.7,circle,fill=color_L52,draw=color_L52Contour](){};}}}

\newcommand{\markerLVaries}{\raisebox{0.5pt}{\tikz{\node[draw,scale=0.5,regular polygon, regular polygon sides=3,fill=none,color=color_LVaries](){};}}}
\newcommand{\markerDVaries}{\raisebox{0.5pt}{\tikz{\node[draw,scale=0.5,regular polygon, regular polygon sides=3,fill=none,color=color_DVaries, rotate = 180](){};}}}
\newcommand{\markerTauVaries}{\raisebox{0.5pt}{\tikz{\node[draw,scale=0.6,regular polygon, regular polygon sides=4,fill=none,color=color_TauVaries,rotate=45](){};}}}
\newcommand{\markerLFiveTwo}{\raisebox{0.5pt}{\tikz{\node[draw,scale=0.7,circle,fill=none,color=color_L52](){};}}}

\newcommand{\markerExpVelocity}{\raisebox{0.5pt}{\tikz{\node[draw,scale=0.5,circle,fill=none,color=color_VelExp](){};}}}
\newcommand{\markerExpEvolution}{\raisebox{0.5pt}{\tikz{\node[draw,scale=0.5,circle,fill=none,color=color_Blue](){};}}}
\newcommand{\markerSimuEvolution}{\raisebox{0.5pt}{\tikz{\node[draw,scale=0.5,circle,fill=color_Blue,color=color_Blue](){};}}}

\newcommand{\markerSimuTwoD}{\raisebox{0.5pt}{\tikz{\node[draw,scale=0.5,regular polygon, regular polygon sides=4,fill=color_Blue,color=color_Blue](){};}}}
\newcommand{\markerSimuAxi}{\raisebox{0.5pt}{\tikz{\node[draw,scale=0.5,circle,fill=color_Red,color=color_Red](){};}}}

\title{Vortex rings generated by a translating disk from start to stop}


\author{Joanne Steiner}
\email[]{joanne.steiner@universite-paris-saclay.fr}
\affiliation{Universit\'e Paris-Saclay, CNRS, Laboratoire FAST, F-91405 Orsay, France}
\author{Cyprien Morize}
\email[]{cyprien.morize@universite-paris-saclay.fr}
\affiliation{Universit\'e Paris-Saclay, CNRS, Laboratoire FAST, F-91405 Orsay, France}
\author{Ivan Delbende}
\affiliation{Sorbonne Universit\'e, CNRS, Institut Jean Le Rond d'Alembert, F-75005 Paris, France}
\author{Alban Sauret}
\affiliation{University of California, Santa Barbara, Department of Mechanical Engineering, CA 93106, USA}

\author{Philippe Gondret}
\affiliation{Universit\'e Paris-Saclay, CNRS, Laboratoire FAST, F-91405 Orsay, France}


\date{\today}

\begin{abstract}
In this article, we investigate experimentally and numerically the time evolution of vortex rings generated by the translation of a rigid disk in a fluid initially at rest and submitted to an acceleration followed by a deceleration. The size of the disk and its motion in terms of stroke length and travel time are varied as control parameters. The start-up vortex ring created in the near wake of the disk is characterized experimentally by PIV, and the measurements agree quantitatively with axisymmetric numerical simulations performed with the Basilisk flow solver. The maximum radius and circulation of the annular vortex and its dynamics are shown to follow different power laws with the control parameters. The modeling adapted from Wedemeyer's two-dimensional theoretical calculations [E. Wedemeyer, Ausbildung eines Wirbelpaares an den Kanten einer Platte, Ingenieur-Archiv 30, (1961)] captures the observed scaling laws. Besides, after the disk stops, a secondary ``stopping" vortex ring is generated, which is shown to affect the motion of the main vortex ring.
\end{abstract}


\maketitle


\section{Introduction} \label{SecI}
The motion of a solid object within a fluid leads to the formation of vortices \cite{Pierce1961, Pullin1980}, a phenomenon linked to the exchange of momentum between solid and fluid. This phenomenon can be observed in steady configurations such as aircraft in cruise flight or rotating wind turbines where, as a reaction to loading, vortices emanating from the wing or blade tips and trailing edges are continuously created in their wakes. Yet, biological systems such as animals use unsteady motions of solid parts to propel and lift themselves, to maneuver, and to hide \cite{Wang2005}. This includes wing flapping, fish swimming, fin waving, etc. The unsteadiness gives rise to a more complex vortex wake system with an array of interconnected rings, loops, and straight parts \cite{Vogel1994}. In the present paper, we investigate the vortex ring created in the wake of a flat disk during a single stroke, from start to stop. The focus of this study is not set on the forces acting on the solid disk but rather on the physical properties of the vortex ring generated: size of the core radius, circulation, and trajectory. The objective is to relate the parameters of the system (disk radius, stroke length, travel time) to the characteristics of the vortices, to be able to assess their impact on the environment. Applications cover biological systems such as the locomotion of animal groups, and the camouflage of flatfish \cite{McKee16}, but also industrial systems using bio-inspired propulsion, naval architecture, and offshore engineering \cite{dabiri2009optimal,schnipper2009vortex,becker2015hydrodynamic,Costa21,menzer2022bio}.  More generally, the understanding of vortex formation is important for the design of new technologies such as wind turbines and energy harvesting systems \cite{whale2000experimental,barrero2012extracting}.

The start-up vortex generated when a flat plate is accelerated from rest is a classical flow problem in fluid dynamics, first initiated by Prandtl in 1924 \cite{Prandtl1924}. It has been established that the sharp edge sets the boundary layer separation point \cite{Prandtl1924}. As the object translates, vorticity is continuously generated under the form of a thin layer advected to the near wake region, where it rolls up to form a vortex. 
Pullin showed that the spiral sheet leads to a quasi-circular vortex in the flat-plate case, elliptical distortions appearing for sheets past a wedge \cite{Pullin1978}. Over the years, much attention has been given to the roll-up of a two-dimensional vortex sheet in an ideal (inviscid) fluid \cite{Wagner1925, Kaden1931, Anton1939, wedemeyer_ausbildung_1961, Pullin1978}.  Wagner proposed a linear theory to account for the two-dimensional starting flow of an ideal fluid in the near wake of a flat plate \cite{Wagner1925}. However, the linear theory is not able to accurately describe the starting vortex over a short period of time. Kaden argued that the transfer of vorticity from the shear layer is responsible for the vortex expansion and demonstrated that the size of the vortex sheet grows in time as $t^{2/3}$ \cite{Kaden1931}. Anton \cite{Anton1939} and Wedemeyer \cite{wedemeyer_ausbildung_1961} proposed to use self-similar solutions as a basis for computing the vorticity shedding in the wake of a moving semi-infinite plate starting impulsively and translating at a uniform velocity. Wedemeyer's results provided more details, including the rate of growth, the shape, and the total circulation of the vortex sheets, which were calculated graphically through step-by-step integration. Although the viscosity of the fluid is responsible for the formation of shear layers and diffusion of vorticity leading to the smoothing of the vortex core, the previous studies mentioned here have disregarded its significance, which appears to be justified for high Reynolds number flows where inertial effects predominate.

Following the theoretical approach of the two-dimensional start-up vortex generation, some studies started to focus on the generation of a vortex ring in an axisymmetric configuration. There are basically two methods to generate vortex rings: by pushing a fluid column out of a circular orifice with a piston or by translating a circular disk in a fluid. The first mechanism has been widely analyzed \cite{maxworthy_experimental_1977, Gharib1998, Dabiri2004, Dabiri2009}. The scaling laws obtained in the previous studies and subsequent refinements \cite{Auerbach1987}, that predict the size, shape, and dynamics of the vortices that form in the wake of a moving plate, have been used in previous research \cite{saffman_1978,didden_formation_1979,hettel_2007}. The second mechanism, on the contrary, has received less attention up to now. Taylor was the first to address theoretically the characteristics of a starting vortex ring produced by a disk in translation \cite{Taylor1952}. By considering that the disk moves in an inviscid fluid without flow separation and suddenly disappears \cite{Lamb}, Taylor showed that the characteristics of the vortex ring, using Lamb's model, are completely determined by the radius $R$ and the velocity $U$ of the disk. He demonstrated that the circulation of the vortex ring is $4UR/\pi$, that the self-induced velocity of the vortex ring is $0.436 U$, and that the vortex ring radius $R_\mathrm{ring}$ and the vortex core radius $a$ are given by $R_\mathrm{ring}=0.816 R$ and $a=0.152 R$ respectively. Comparing this prediction with experimental data is, however, a delicate issue since Taylor's derivation can only be used at the very beginning of the vortex formation \cite{Sallet1975}. 

Numerical simulations conducted by Shenoy and Kleinstreuer \cite{Shenoy2008} showed that the Reynolds number can cause the flow of a vortex ring produced by a translating disk to switch from an axisymmetric to a three-dimensional periodic state. More recently, Yang et al \cite{Yang2012} investigated the dynamics of vortex ring formation behind a circular disk. They identified three stages for this process. Initially, there is a rapid growth of the vortex circulation up to $tU/2R < 0.2$, where Taylor's inviscid estimation is accurate. This is followed by a phase of stable growth where the rate of growth of the circulation decreases gradually. After $tU/2R > 4$, the vortex ring loses its axisymmetry due to instabilities. However, the effects of the radius, velocity, and stroke length of an impulsively started circular disk on the characteristics of the resulting vortex ring are still elusive.

The present paper focuses on the generation of a vortex ring in the near wake of a disk in translation, with a monotonic sinusoidal motion from start to stop. The experimental setup and the numerical methods for investigating this configuration are presented in section II. The experimental and numerical results concerning the main features of the vortex rings (circulation, radius, position) as a function of the control parameters (diameter, stroke length, and travel time of the disk) are then detailed in section III. These results are then discussed and rationalized in section IV where we provide scaling laws based on the theoretical frame of Wedemeyer \cite{wedemeyer_ausbildung_1961} adapted to the present axisymmetric disk configuration and where we discuss the behavior of the vortex ring after the disk stops. A conclusion for these experimental, numerical, and modeling approaches is finally provided in section V. \\

\section{Methods} \label{SecII}

\subsection{Experimental setup}

\begin{figure}[t]
     \centering
     \includegraphics[width=0.9\linewidth]{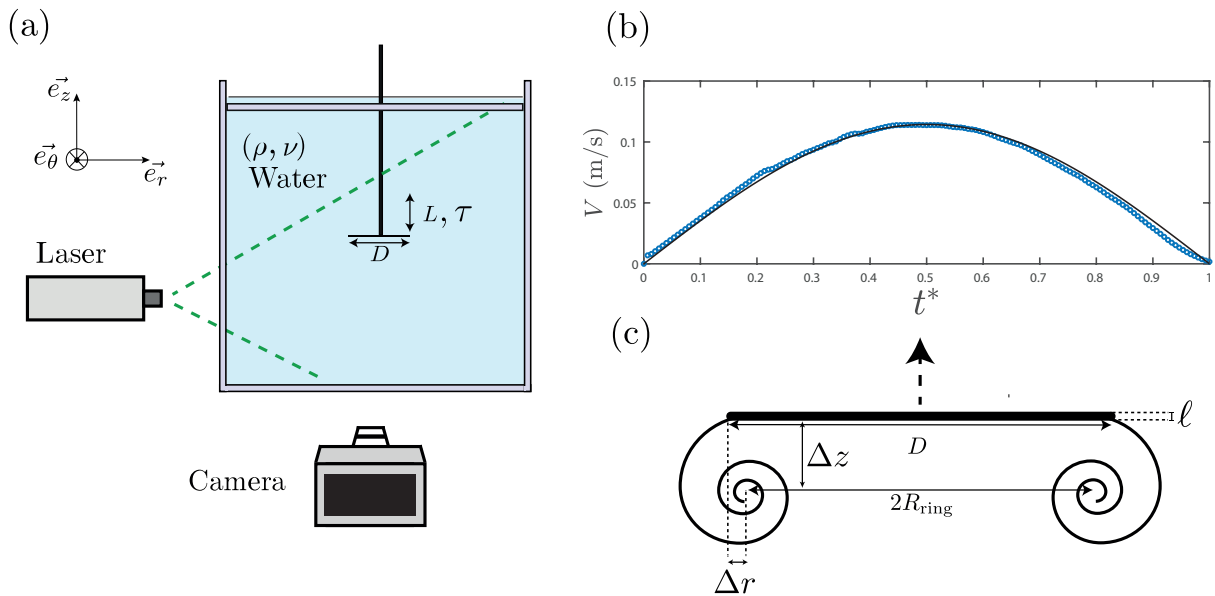}
        \caption{(a) Sketch of the experimental setup. (b) Time evolution of the velocity of the disk for $L$ = 5.2 cm and $\tau = 0.71~$s ($D = 10~$cm). (\protect\markerExpVelocity) \ Experimental velocity and (---) expected velocity given by equation (\ref{eq:motion_law}). (c) Definition of the main properties of the vortex ring in the near wake of the translating disk.}
        \label{fig:IIA_SetUpExpAndVortex}
\end{figure}

A schematic of the experimental setup is shown in figure \ref{fig:IIA_SetUpExpAndVortex}(a). Experiments are performed in a tank of square cross-section 40 cm $\times$ 40 cm and of height 60 cm. The tank is filled with water at ambient temperature over a height of 40 cm. A rigid disk of diameter $D$ and parallel to the bottom of the tank is set into vertical translation along the $z$ direction. The disk is placed at the center of the tank at a distance from the walls as large as possible to be in a quasi-unbounded flow. We ensured that slightly changing the initial position of the disk does not modify the results presented in the following. The diameter $D$ of the disk is varied between 5 cm and 15 cm, and its thickness $\ell$ = 2 mm is kept constant so that $\ell \ll D$. The vertical translation of the disk is performed by an AC servo motor (ECMA-C20807RS) and an eccentric system that are not represented in figure \ref{fig:IIA_SetUpExpAndVortex}(a). The eccentric system converts the rotational motion of the motor into a translation of the disk. The stroke length $L$ can be adjusted  between 2 cm and 6 cm by changing the eccentric settings, meaning that it is at least ten times larger than the disk thickness $\ell$. Indeed, the stroke length $L$ must be large enough compared to the disk thickness $\ell$ to generate a vortex ring that can be experimentally resolved.

The motor performs half a rotation, making the disk move in one direction, upward or downward, with a sinusoidal acceleration and deceleration. The travel time $\tau$ corresponds to the time taken by the disk to travel the stroke length $L$ and is varied between 0.25 s and 2.5 s. A rigid plate placed just below the free surface avoids the presence of surface waves. 

In the following, we define the non-dimensional time as $t^* = t/\tau$. The motion of the disk starts at $t^* = 0$ and ends at $t^* = 1$, and the instantaneous velocity of the disk for $0 \leq t^* \leq 1$ is given by:
\begin{equation}
V(t^*) = V_\text{m} \sin(\pi t^*),
\label{eq:motion_law}
\end{equation}
where $V_\text{m} = \pi L/(2\tau)$ is the maximum velocity of the disk reached at mid-stroke. Note that the velocity is always positive, which means that the disk goes up. All the results shown in the following were obtained for a disk going up and are transferable for a disk going down. An example of the time-evolution of the velocity of the disk, for $L = 5.2~$cm and $\tau$ = 0.71 s ($D$ = 10 cm) is reported in figure \ref{fig:IIA_SetUpExpAndVortex}(b) and shows that the experimental and the prescribed velocity of the disk agree well. In the present flow configuration, the relevant dimensionless numbers are the Reynolds number based on the disk diameter $Re$ = $V_\text{m} D/\nu$, where $\nu = 10^{-6}$ m$^2$/s is the kinematic viscosity of water at room temperature, and the reduced Keulegan--Carpenter number which here corresponds to the relative stroke length $L/D$. These non-dimensional numbers vary in the range $Re \in [10^3, ~2\times 10^4]$ and $L/D \in [0.2,~ 0.83]$. Note that the Keulegan-Carpenter number $L/D$ is small compared to 4, meaning that the vortex formation is expected to be mainly axisymmetric \cite{Yang2012}.

The flow field is characterized experimentally by Particle Image Velocimetry (PIV) measurements. A Powell lens, placed right after a continuous 2W Nd-Yag laser, transforms the beam into a vertical sheet that illuminates a plane passing through the axis of the disk. The water is seeded with hollow glass particles Sphericel\textsuperscript{\textregistered} 110P8 of median diameter 10 $\mu$m and of density 1.1 g/cm$^3$ close to the density of water. The experiments are recorded with a high-speed camera (Phantom MIRO M110) equipped with a 85 mm lens. The frame rate of the camera is adapted for each experiment and ranges between 40 fps to 400 fps. The computations of the velocity fields are performed with the software DAVIS (LaVision) \cite{Davis}. The size of the interrogation windows is between 16$\times$16 pixels$^2$ and 24$\times$24 pixels$^2$ with an overlap of 75$\%$. Since the flow is recorded in a vertical plane centered in the meridional plane of the disk, each individual realization shows some minor differences due to slight non-axisymmetric fluctuations in the flow. Hence, convergence is achieved by computing the average of 20 independent realizations. The coordinate system ($r$, $\theta$, $z$), represented in figure \ref{fig:IIA_SetUpExpAndVortex}(a), is centered on the disk, and $z = 0$ corresponds to the starting position of the disk. Therefore, the disk lies between $r = 0$ and $r = D/2$ for any $\theta \in [0 , 2 \pi]$ and its vertical position at $t^*$ is $z_\mathrm{d}(t^*) = L[1-\cos(\pi t^*)]/2$ so that $z_\mathrm{d}(0) = 0$ and $z_\mathrm{d}(1) = L$. Note that the velocity field $(v_r, v_z)$ is recorded only for $\theta = 0$ since the flow is assumed to be mainly axisymmetric.

A vortex ring forms in the wake of the translating disk, and the resulting PIV measurements are analyzed with MATLAB custom-made routines \cite{Matlab}. As sketched in figure \ref{fig:IIA_SetUpExpAndVortex}(c), we extract the horizontal and vertical position of the vortex relative to the disk, $\Delta r$ and $\Delta z$, respectively, from the velocity field using the second-moment method detailed in section \ref{section:MethodVortex}. The radius of the vortex ring $R_\text{ring}$ can be deduced from $\Delta r$ using the following relation: $R_\text{ring} = D/2-\Delta r$. The circulation $\Gamma$ and the core radius $a$ of the vortex are also extracted at each time step.

\subsection{Numerical method}
\begin{figure}[t]
    \centering
    \includegraphics[width=\linewidth]{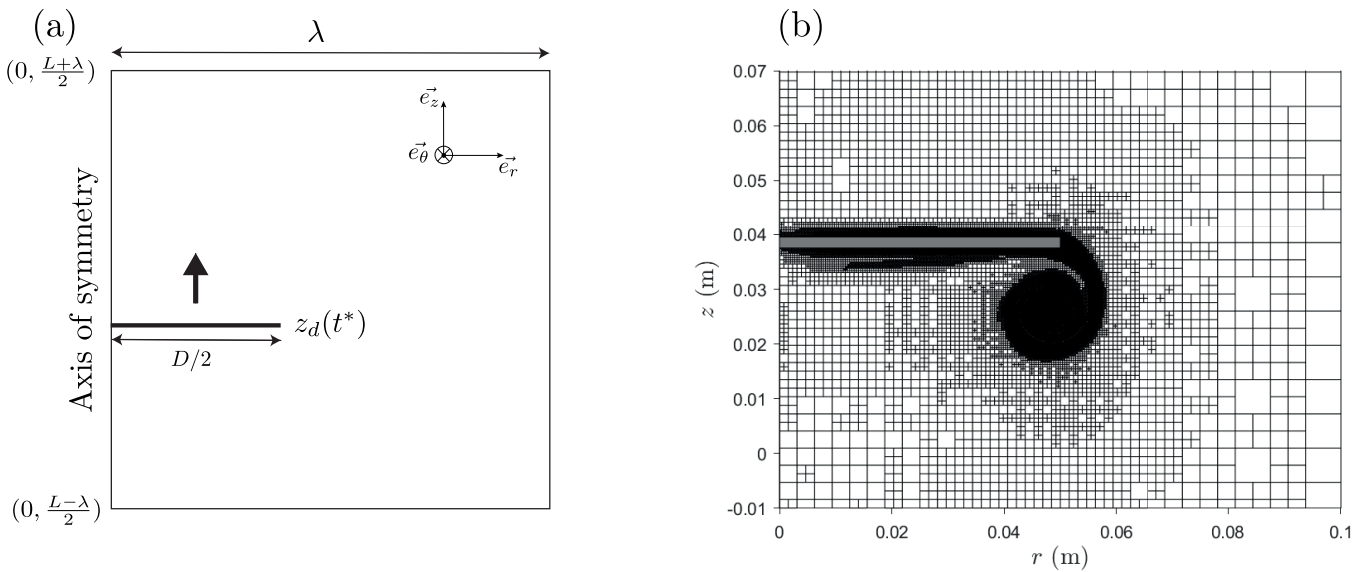}
    \caption{(a) Sketch of the numerical setup. The disk is going up and is initially placed at the vertical position $z$ = 0. (b) Example of the adaptative mesh refinement around the disk for $L$ = 5.2 cm, $D$ = 10 cm, and $\tau$ = 1 s at $t^*$ = 0.66. }
    \label{fig:SetUpNum}
\end{figure}

The flow field generated by the translation of the disk is also determined using direct numerical simulations of the Navier--Stokes equations for a Newtonian incompressible fluid using the Basilisk flow solver \cite{Basilisk}. A sketch of the computational domain is given in figure \ref{fig:SetUpNum}(a). The configuration is 2D axisymmetric: in the meridional plane, the computational domain is a square of side $\lambda$ defined by $(r,z)\in [0,\lambda]\times[(L-\lambda)/2,(L+\lambda)/2]$, where $\lambda=4D$. Modifying slightly the domain size $\lambda$ does not alter the results. The solid disk is taken into account through an immersed boundary. At $t=0$, the disk located within the region $[0,D/2]\times[-\ell/2,\ell/2]$, is represented by a solid volume fraction. This region is then moved in time at the velocity prescribed by equation (\ref{eq:motion_law}), this latter velocity being enforced to the region containing the disk using a solid volume fraction. The left boundary has an axisymmetric boundary condition. At the outer boundaries (top, bottom, and right), no-slip conditions are used.

The numerical scheme uses cell-centered velocity/pressure $(v_r, v_z, p)$ variables and involves an explicit upwind Bell--Collela--Glaz advection scheme, while viscous terms are treated implicitly. The spatial discretization is based on a regular Cartesian mesh, with an adaptive refinement through a quadtree approach \cite{Popinet2009}.
More specifically, the domain is initially a uniform grid. The adaptive algorithm computes the numerical errors on the values of $v_r$ or $v_z$ for each square cell. Depending on its numerical error, each cell is coarsened, refined or kept the same. A typical example of the mesh grid is shown in figure \ref{fig:SetUpNum}(b). The adaptive algorithm enhances precision in regions that need so, namely at solid boundaries and near high velocity gradient zones. It also drastically reduces the cost of computations compared to the case where the maximum refinement level would be enforced in the whole domain. Here, we have ensured that the size of the finest cells is at least ten times smaller than the thickness $\ell$ of the disk so that the mesh is fine enough at the edge of the disk. A vorticity sheet, responsible for the growth of the vortex, is generated in the vicinity of the moving disk. The number of cells in this vorticity sheet is around 10. The time-step is chosen by imposing a Courant number $\text{CFL} = 0.8$ and a maximum time-step $\delta t_\text{max}$ = $0.1\tau$. Hence, the time-step $\delta t$ is computed as $\delta t = \min(\text{CFL} |\Delta/v|_\text{min}, \delta t_\text{max})$, where $\Delta$ is the size of a cell and $v$ the radial or vertical velocity in this cell.

We ensured that a higher refinement level changes the circulation of the vortex ring by less than 1.4\% and that results are unaffected by changing the thresholds used as refinement criteria, in the vicinity of the values adopted here. 

In the simulations, the range of stroke lengths and diameters of the disk has been extended so that $L \in [2, 20]~$cm and $D\in[5, 40]~$cm and the non-dimensional numbers vary in the range $Re \in [10^3, ~2.6\times 10^4]$ and $L/D \in [0.07,~ 2]$.

\subsection{Properties of the vortex}
\label{section:MethodVortex}
The properties of the vortex are extracted from the vorticity field using the second-moment method \cite{le_dizes_viscous_2002}. This method is particularly useful when vortices are elliptical and not aligned with the main axes. In addition to the position $(r_\mathrm{G}, z_\mathrm{G})$ of the vortex and its circulation $\Gamma$, this method allows to evaluate two radii $(a_\mathrm{r}, a_\mathrm{z})$ and the orientation angle $\alpha_\mathrm{G}$ of the major axis of the ellipse with respect to the $r$-axis, as explained in the following. First, the position of the vortex is found by computing the coordinates of the barycenter of the vorticity $\omega$:
    \begin{equation}
        r_\mathrm{G} = \iint_S \frac{r\, \omega}{\Gamma} ~\mathrm{d} r \, \mathrm{d}z,\quad   
        z_\mathrm{G} = \iint_S \frac{z\, \omega}{\Gamma} ~\mathrm{d}r\, \mathrm{d}z,
    \end{equation}
    where $\omega = \partial v_r/\partial z - \partial v_z/\partial r $ and $\Gamma$ is the total circulation  given by
    \begin{equation}
        \Gamma = \iint_S \omega ~\mathrm{d}r \, \mathrm{d}z.
    \end{equation}
    The circulation is computed over a domain of surface $S$ large enough to enclose most of the vorticity associated to the vortex.
    The angle $\alpha_\mathrm{G}$ between the $r$-axis and the vortex major axis is solution of
    \begin{equation}
        \iint_S \bigg[r^{(\alpha_\mathrm{G})}-r_\mathrm{G}^{(\alpha_\mathrm{G})}\bigg]\bigg[z^{(\alpha_\mathrm{G})}-z_\mathrm{G}^{(\alpha_\mathrm{G})}\bigg] ~\omega ~\mathrm{d}r \, \mathrm{d}z = 0,
    \end{equation}
     where $r^{(\alpha_\mathrm{G})}$ and $z^{(\alpha_\mathrm{G})}$ are coordinates along axes obtained through rotation of the $r$ and $z$-axes by the angle $\alpha_\mathrm{G}$:
    \begin{equation}
        r^{(\alpha_\mathrm{G})} = r\cos \alpha_\mathrm{G} + z \sin \alpha_\mathrm{G} ,\quad
        z^{(\alpha_\mathrm{G})} = z\cos \alpha_\mathrm{G} - r \sin \alpha_\mathrm{G} .
    \end{equation}
    The radii $(a_\mathrm{r},a_\mathrm{z})$ of the elliptical vortex  along the $r^{(\alpha_\mathrm{G})}$ and $z^{(\alpha_\mathrm{G})}$-axes are such that
    \begin{equation}
        {a_\mathrm{r}}^2 = \iint_S \frac{[r^{(\alpha_\mathrm{G})}-r_\mathrm{G}^{(\alpha_\mathrm{G})}]^2}{\Gamma}~\omega ~\mathrm{d}r \, \mathrm{d}z,\ \ \ 
        {a_\mathrm{z}}^2 = \iint_S \frac{[z^{(\alpha_\mathrm{G})}-z_\mathrm{G}^{(\alpha_\mathrm{G})}]^2}{\Gamma}~\omega ~\mathrm{d}r \, \mathrm{d}z.
    \end{equation}
Finally, the dispersion radius $a$ of the vortex core \cite{le_dizes_viscous_2002} and its ellipticity $\varepsilon$ are defined as
    \begin{equation}
        a = \sqrt{{a_\mathrm{r}}^2+{a_\mathrm{z}}^2},\ \ \ \varepsilon = \max \left (\frac{a_\mathrm{r}}{a_\mathrm{z}},\frac{a_\mathrm{z}}{a_\mathrm{r}} \right ),
        \label{eq:ellipticity}
    \end{equation}
respectively. An iterative routine based on these equations is implemented in Matlab. The total circulation of the vortex is computed on a centered surface of radius five times bigger than the radius of the vortex to capture the major part of its vorticity. However, vorticity of opposite sign is excluded from the calculation to prevent the inclusion of the vorticity concentrated into the boundary layer near the disk.

\section{Results} \label{SecIII}
\subsection{Phenomenology}

\begin{figure}
    \centering
    \includegraphics[width=0.75\linewidth]{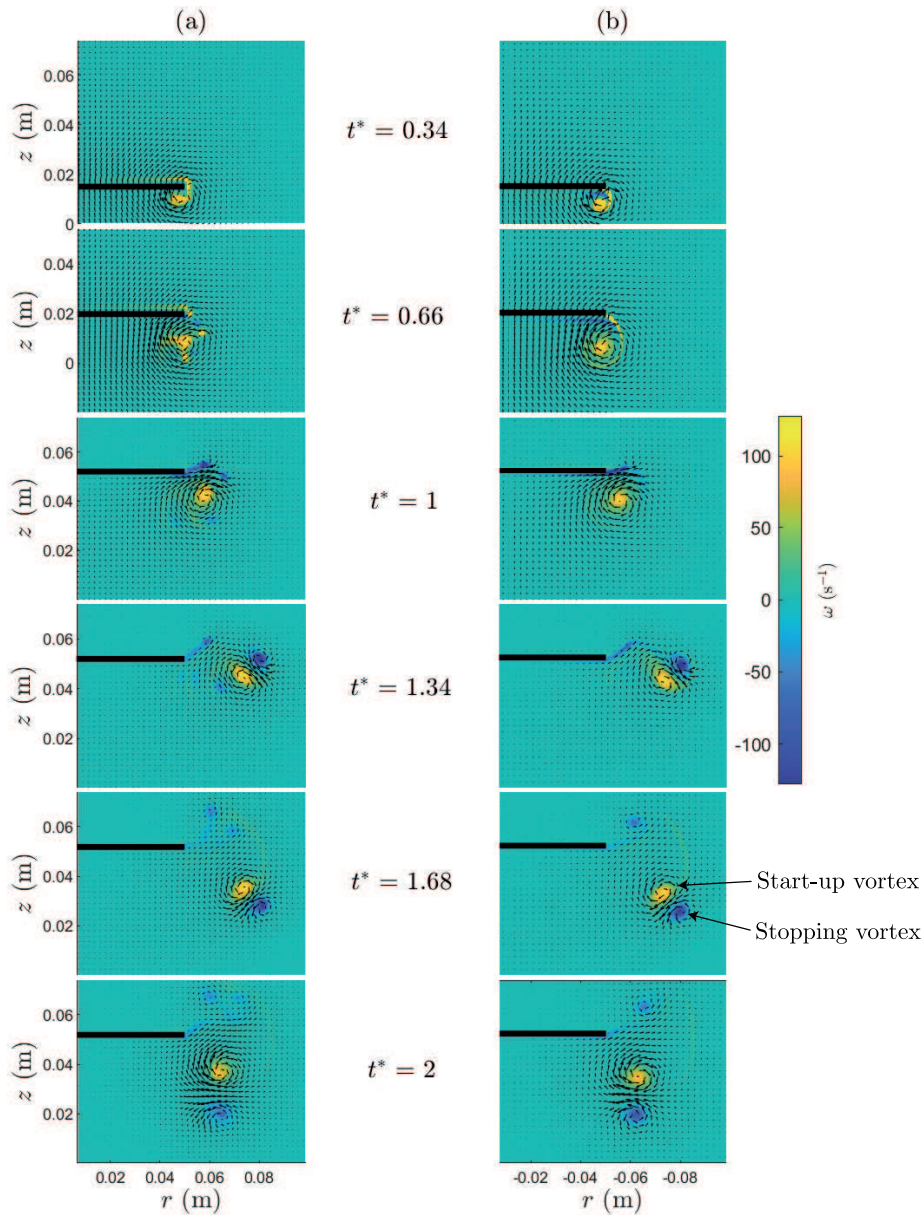}
    \caption{Snapshots of the velocity fields (arrows) and the vorticity fields (color scale) generated by the upward translation of the disk in (a) the experiments and (b) the numerical simulations at different non-dimensional time $t^*$ for $L$ = 5.2 cm, $D = 10~$cm, $\tau$ = 1 s.}
    \label{fig:III_VorticitySnapshots}
\end{figure}

We describe in this section the structure of the flow generated by the translating disk. In figures \ref{fig:III_VorticitySnapshots}(a)-(b), snapshots of the velocity field (arrows) and vorticity fields (color scale) are reported for an experiment and for the corresponding numerical simulation respectively at different times for $L$ = 5.2 cm, $D$ = 10 cm and $\tau$ = 1 s. The translation of the disk generates a vortex in its near wake, and the experimental and numerical behaviors of the vortex ring are similar. During the motion of the disk, for $t^*\leq 1$, circulation is enrolled in a start-up vortex, making its radius grows and the maximum vorticity increases. At $t^*$ = 0.66, the vortex in the numerical simulation exhibits a well-defined tail of vorticity that connects the vortex to the edge of the disk. When computing the size of the vortex, as described in section \ref{section:MethodVortex}, the tail of vorticity  that goes from the edge of the disk to the vortex is not included so that the radius obtained corresponds only to the core of the vortex. However, the tail of vorticity is kept for the computation of the position of the vortex and its circulation since it does not significantly influence these quantities. Moreover, we perform a comparison of the circulation of the vortex with a theoretical approach of a start-up vortex flow \cite{wedemeyer_ausbildung_1961} that includes all the vorticity sheet that has rolled up and consequently the tail of the vortex. 

In the experiments, small satellite vortices are observed around the main vortex (see, for instance, figure \ref{fig:III_VorticitySnapshots} at $t^*$ = 0.66). The formation of these irregularities in a vorticity sheet has been previously observed in numerical simulations \cite{Luchini2002,Luchini2017, higuchi_numerical_1996} and experiments \cite{Pierce1961,higuchi_numerical_1996} of the start-up vortex flow of a flat plate and is due to a Kelvin-Helmholtz instability. At this Reynolds number, the instability cannot be observed in the numerical simulations. At $t^* = 0.66$, both in the experiment and the simulation, a thin layer of opposite vorticity is created between the disk and the vortex ring.

When the disk stops, at $t^*$ = 1, the vorticity tail detaches from the edge of the disk, and no more circulation is enrolled in the vortex ring. At the same time, the roll-up velocity of the vortex induces the generation of a secondary or “stopping” vortex of opposed circulation at the edge of the disk. After the disk has stopped ($t^* > 1$), the primary vortex ring is completely detached from the disk and evolves while interacting with the secondary vortex of opposite circulation. From $t^* = 1$ to $t^* = 2$, the vortices move and rotate under their mutual interaction. At $t^* = 1.34$ and $t^* = 1.68$, it can be seen that the main vortex is also deformed by the strain field induced by the stopping vortex and becomes elliptical. At $t^*$ = 2, the main vortex and the secondary vortex are almost circular, and their maximum absolute vorticity has reduced. They interact less and gradually diffuse in the fluid due to viscous effects.

\begin{figure}[t]
    \centering
    \includegraphics[width=\linewidth]{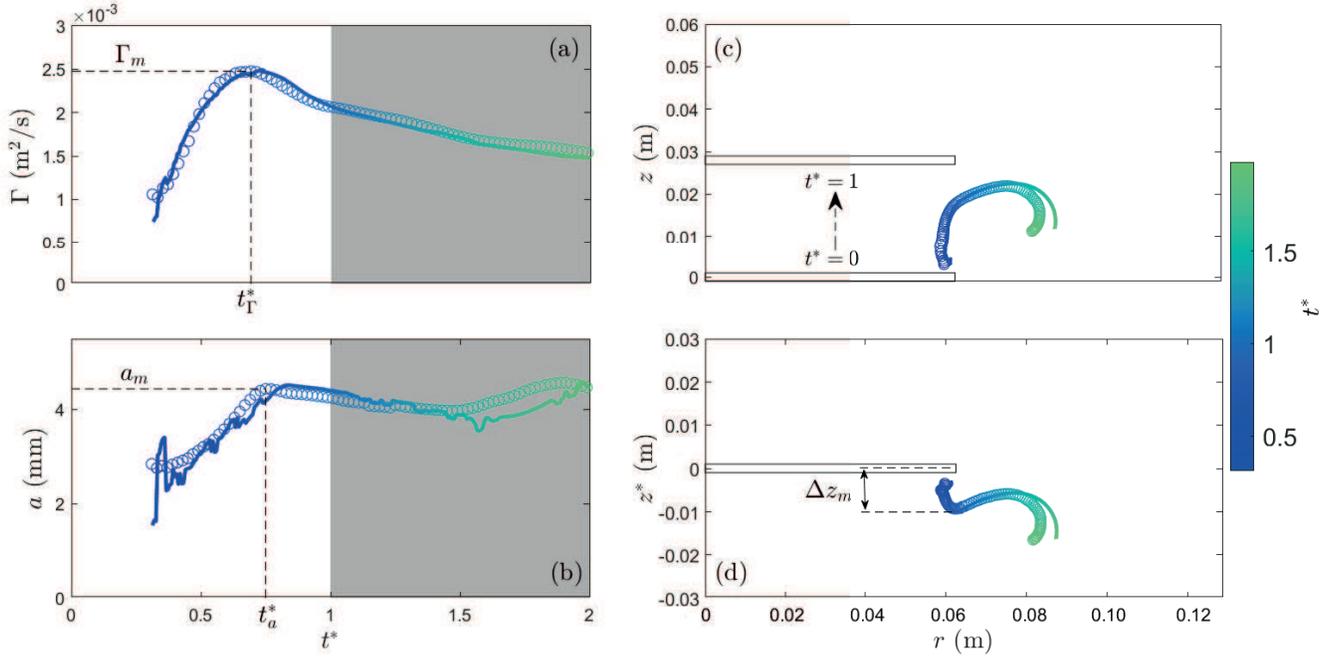}
    \caption{Time evolution of the characteristics of the start-up vortex in the same configuration than in figure \ref{fig:III_VorticitySnapshots} (for $L$ = 2.8 cm, $D = 12.5~$cm and $\tau$ = 1.67 s) for the experiment (\protect\markerExpEvolution) and for the simulation (\textcolor{blue}{\textbf{-}}). (a) Circulation $\Gamma$, (b) radius $a$, and position in the frame of reference (c) of the laboratory and (d) of the disk. For ease of visualization, only one point every five time-steps is displayed for the experimental results.}
    \label{fig:CompExpSimu}
\end{figure}

To describe more quantitatively the behavior of the start-up vortex, the time-evolution of its circulation $\Gamma$ and of its core radius $a$ are reported in figures \ref{fig:CompExpSimu}(a) and \ref{fig:CompExpSimu}(b), respectively, for $L$ = 2.8 cm, $D$ = 12.5 cm and $\tau = 1.67~$s. The position of the core of the vortex is reported in the reference frame of the laboratory in figure \ref{fig:CompExpSimu}(c) and in the reference frame of the disk in figure \ref{fig:CompExpSimu}(d). The origin of the axis $z^*$ in figure \ref{fig:CompExpSimu}(d) corresponds to the position at time $t^*$ of the disk $z_\mathrm{d}(t^*)$ : $z^* = z-z_\mathrm{d}(t^*)$. For each figure, the experimental results (empty symbols) are compared with the corresponding numerical results (solid lines). The results are shown starting at $t^*$ = 0.3 since before this time the vortex is too small to be properly characterized by the routine described in section \ref{section:MethodVortex}. Before describing the time evolution of the physical properties of the start-up vortex ring, we can note that the different properties computed from the experiment and from the numerical simulation agree quantitatively well. Nevertheless, we observe a discrepancy for the vortex radius $a$ at short times in figure \ref{fig:CompExpSimu} (b). Indeed, the calculation of the vortex radius requires a sensitive process of extracting the vorticity tail from the numerical vorticity fields, which in turn generates the observed measurement noise.

We observe in figures \ref{fig:CompExpSimu}(a) and \ref{fig:CompExpSimu}(b) that the time evolution of the vortex can be decomposed into three stages. The first phase corresponds to the generation of the primary vortex. The vortex grows in size and circulation and reaches a maximum radius $a_m$ and circulation $\Gamma_m$  at time $t^*_a \simeq 0.75 $ and $t^*_{\Gamma} \simeq 0.7$, respectively, as indicated in figures \ref{fig:CompExpSimu}(a)-(b). The centroid of the vortex ring also moves away from the disk vertically while its radial position is almost constant when $t^* < 1$ as can be seen in figures \ref{fig:CompExpSimu}(c) and \ref{fig:CompExpSimu}(d). In the following, the maximum vertical distance between the disk and the centroid of the vortex ring before the disk stops is noted $\Delta z_m$ and is reached at time $t^*_{\Delta z}$. In the example of figure 4(d), we find $t^*_{\Delta z} \simeq 0.87$. The time of maximum circulation, radius, and vertical position vary slightly for different sets of parameters. The first stage ends when the circulation starts to decrease. For the present set of parameters, the first phase ends at $t^* \simeq 0.7$, but this value can vary slightly for different sets.

In the second phase, from $t^* \simeq$ 0.7 to the end of the translation of the disk at $t^* = 1$, the vortex approaches the disk, which is decelerating, and starts to move radially outwards. The circulation and the core radius of the vortex ring decrease.

Finally, during the third stage, after the disk has stopped ($t^* > 1$), the start-up vortex has detached from the disk and its motion is no longer forced by the translation of the disk. The vortex moves away radially from the disk at the beginning of this phase and its circulation and radius gradually decrease. The snapshots in figure \ref{fig:III_VorticitySnapshots} have shown the formation of a stopping vortex which makes the starting vortex ring rotate in the vicinity of the disk as shown in figure \ref{fig:CompExpSimu}(c) and figure \ref{fig:CompExpSimu}(d).

\subsection{Features of the starting vortex ring}
The temporal evolution of the primary vortex has been decomposed into three phases: generation (from $t^* = 0$ to $t^* \simeq 0.7$ for the set of parameters presented previously), decrease (from $t^* \simeq 0.7$ to $t^* \simeq 1$) and detachment from the disk (for $t^* \gtrsim 1$). During the translation of the disk, the circulation, the radius, and the distance of the vortex ring from the disk reach maximum values noted $\Gamma_m$, $a_m$, and $\Delta z_m$, respectively. To better characterize the generation of the start-up vortex, the stroke length $L$, the diameter $D$, and the travel time $\tau$ of the disk have been varied independently. The physical parameters and the non-dimensional numbers, as well as the symbols used for the experimental and numerical data in the following figures are summarized in table \ref{tab:IndividualSymbols}.

\begin{table}[t]
\centering
\setlength{\tabcolsep}{12pt}
\begin{tabular}{ c c c c c c c} 
\multicolumn{7}{c}{}\\
\hline \hline
$L$ (cm) & $D$ (cm) & $\tau$ (s) & $L/D$ & $Re$ & Experiments & Simulations\\
\hline \hline
2-6 & 10 & 1.67 & 0.2-0.6 & $2\times 10^3$- $5.6\times 10^3$ & \markerLVaries & \\
2-20 & 10 & 1.67 & 0.2-2 & $2 \times 10^3$- $2 \times 10^4$ & & \markerLVariesSimu \\
\hline
2.8 & 3.75-15 & 1.67 & 0.19-0.75 & $10^3$- $4\times 10^3$  & \markerDVaries & \\
2.8 & 5-40 & 1.67 & 0.07-0.56 & $10^3$- $10^4$ & & \markerDVariesSimu\\
\hline
2.8 & 10 & 0.25-2.5 & 0.28 & $2\times 10^3$- $2\times 10^4$ & \markerTauVaries & \\
2.8 & 10 & 0.25-2.5 & 0.28 & $2 \times 10^3$- $2 \times 10^4$ & &  \markerTauVariesSimu\\
\hline \hline 
\end{tabular}
\caption{Sets of the experimental and numerical parameters and non-dimensional numbers used in this study with the corresponding data symbols used in figures.}
\label{tab:IndividualSymbols}
\end{table}

The maximum radial distance between the disk and the vortex ring will not be given because, as seen in figure \ref{fig:CompExpSimu}, the radial distance does not vary much during the motion of the disk and its maximum would mainly be the consequence of small computing noises. This point will be further discussed in the following section.

\begin{figure}[b]
    \centering
    \includegraphics[width=\linewidth]{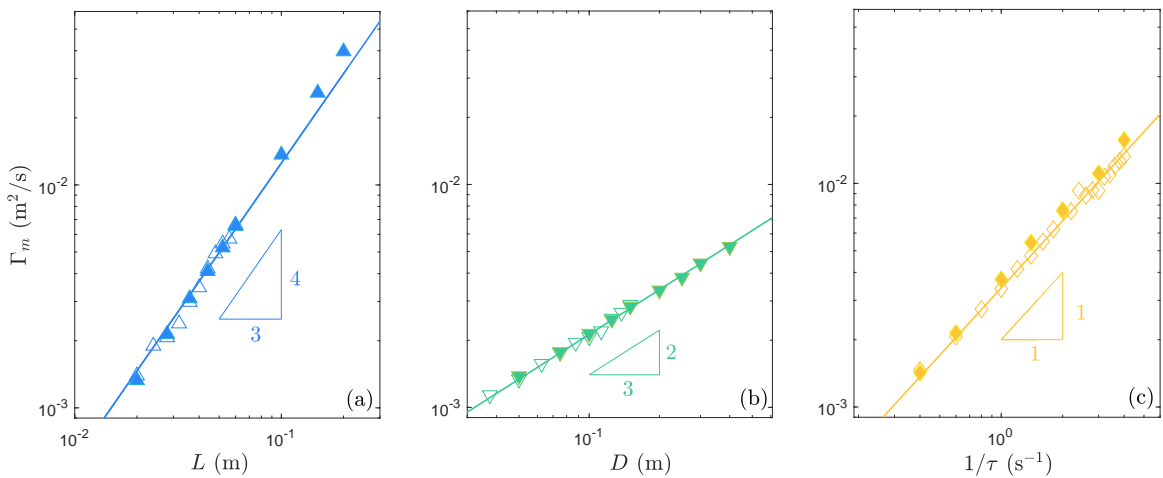}
    \caption{Maximum circulation of the primary vortex $\Gamma_m$ as a function of (a) the stroke length $L$, (b) the diameter $D$, and (c) the inverse of the travel time of the disk $1/\tau$. The parameters kept constant are (a) $D$ = 10 cm, $\tau$ = 1.67s, (b) $L$ = 2.8 cm, $\tau = 1.67~$s, and (c) $L$ = 2.8 cm, $D$ = 10 cm. The lines correspond to fitting by power laws of the experimental (empty symbols) and numerical results (full symbols) and are : (a) $\Gamma_m = \alpha L^{4/3}$ where $\alpha \simeq 0.27$ m$^{2/3}$/s, (b) $\Gamma_m = \beta D^{2/3}$ where $\beta \simeq 0.01~$m$^{4/3}$/s, and (c) $\Gamma_m$ = $\chi /\tau$ where $\chi \simeq 0.0034$ m$^2$.}
    \label{fig:III_ScalingCirculation}
\end{figure}

\begin{figure}[t]
    \centering
    \includegraphics[width=\linewidth]
    {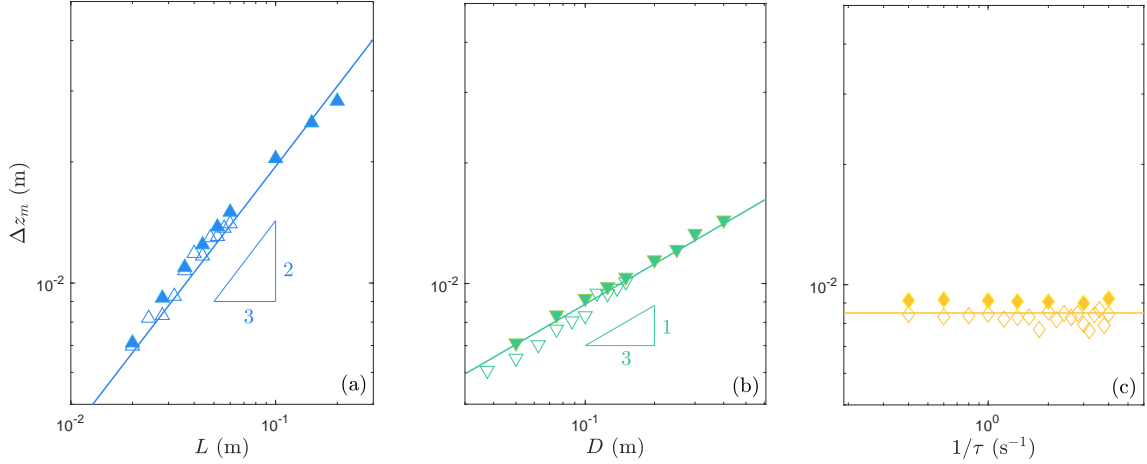}
    \caption{Maximum vertical distance of the vortex $\Delta z_m$ as a function of the same parameters as in figures \ref{fig:III_ScalingCirculation}(a)-(c). The lines are power laws fit of equations : (a) $\Delta z_m$ = $ \alpha L^{2/3}$ where $\alpha \simeq 0.09$ m$^{1/3}$, (b) $\Delta z_m = \beta D^{1/3}$ where $\beta \simeq 0.02$ m$^{2/3}$, and (c) $\Delta z_m \simeq$ 8.5 mm.}
    \label{fig:III_ScalingDeltaz}
\end{figure}

First, the maximum circulation, $\Gamma_m$, is given in figures \ref{fig:III_ScalingCirculation}(a)-(c) as a function of the control parameters $L$, $D$, and $\tau$, for the experiments (empty symbols) and the numerical simulations (full symbols). We observe a quantitative agreement between the numerical simulations and the experiments. The maximum circulation of the vortex ring increases with the stroke length $L$, with the diameter of the disk $D$ and the inverse of the travel time $1/\tau$, \textit{i.e.} with the velocity of the disk. In the range of parameters considered here, power laws capture the evolution of the maximum circulation. More specifically, for the range of stroke length $L$ considered in figure \ref{fig:III_ScalingCirculation}(a) and $D = 10~$cm, $\tau = 1.67 s$, $\Gamma_m$ varies as $L^{4/3}$. For the range of diameter $D$ of the disk considered here and for $L$ = 2.8 cm and $\tau$ = 1.67 s, $\Gamma_m$ varies as $D^{2/3}$. Finally, as shown in figure \ref{fig:III_ScalingCirculation}(c), for the range of travel time $\tau$ considered here and for $L = 2.8$ cm and $D$ = 10 cm, $\Gamma_m$ is proportional to $1/\tau$. 

Secondly, in figures \ref{fig:III_ScalingDeltaz}(a)-(c) the maximum vertical position of the vortex $\Delta z_m$ is shown, for the experiments and the numerical simulations, which are again in excellent quantitative agreement. The vertical position $\Delta z_m$ increases with the stroke length $L$ and the diameter $D$ of the disk but seems independent of the travel time $\tau$. Similarly to the circulation, power laws emerge from the data. For the range of stroke length $L$, diameter $D$, and travel time $\tau$ considered in figures \ref{fig:III_ScalingDeltaz}(a)-(c), $\Delta z_m$ varies as $L^{2/3}$ and $D^{1/3}$.

\begin{figure}[b]
    \centering
    \includegraphics[width=\linewidth]
    {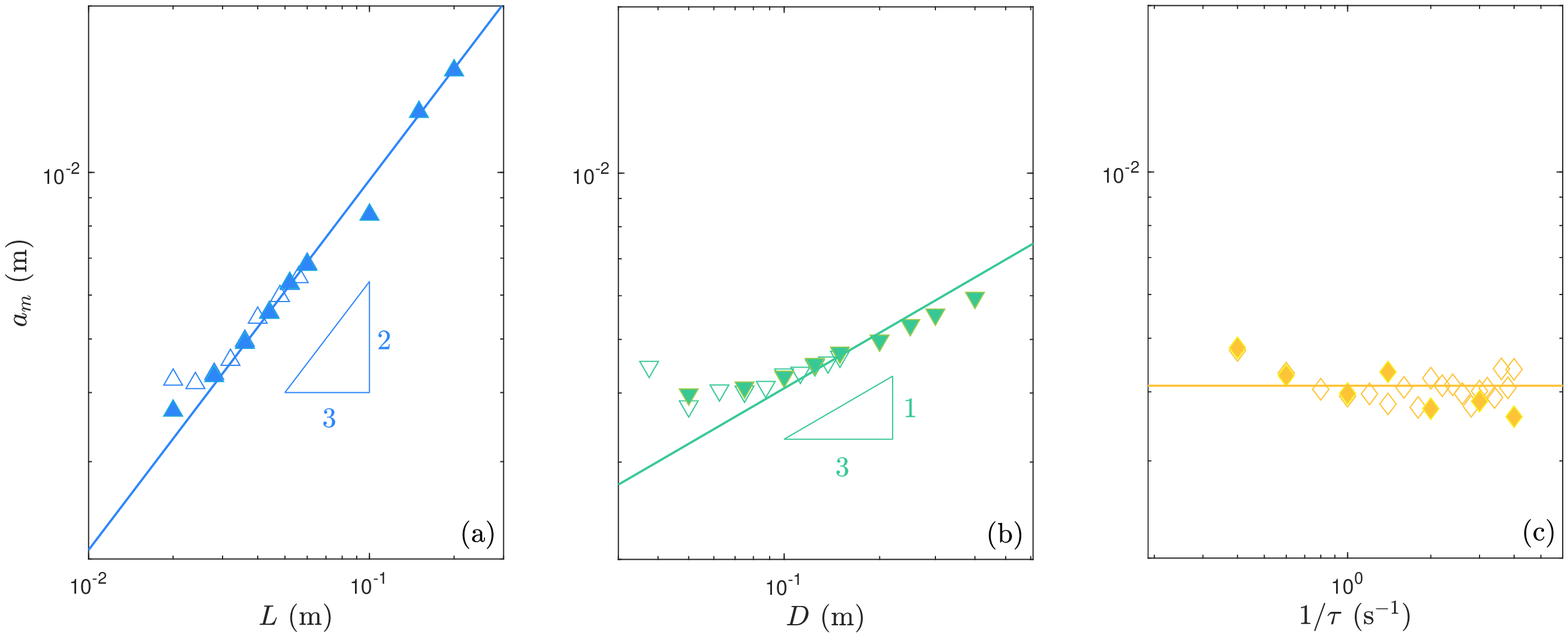}
    \caption{Maximum core radius of the vortex $a_m$ as a function of the same parameters as in figures \ref{fig:III_ScalingCirculation}(a)-(c). The lines are power laws fit of equations : (a) $a_m$ = $\alpha L^{2/3}$ where $\alpha \simeq 0.045$ m$^{1/3}$; (b) $a_m$ = $\beta D^{1/3}$ where $\beta \simeq 0.01$ m$^{2/3}$ and (c) $a_m \simeq 4.1$ mm}
    \label{fig:III_ScalingRadius}
\end{figure}

Finally, the maximum radius of the vortex $a_m$ is given in figures \ref{fig:III_ScalingRadius}(a)-(c). Similar conclusions can be drawn. More specifically, the radius $a_m$ increases with the stroke length $L$ and the diameter $D$ of the disk and is independent of the travel time $\tau$. Scaling laws capture the evolution of $a_m$ and, for the range of parameters considered here, $a_m$ evolves as $L^{2/3}$ and $D^{1/3}$, although there is a slight departure from the power law for the smallest diameter. Indeed the size of the radius of the vortex starts to be non-negligible when the disk diameter becomes too small. The maximum radius of the vortex is found to be proportional to its maximal vertical position, $a_m \simeq 0.48~\Delta z_m$. The radius and the vertical position of the vortex ring are linked because the more the vortex grows, the more it moves away from the disk. Thus, the maximum radius has the same dependency with the control parameters as $\Delta z_m$. 

\section{Discussion}

\subsection{First phase: Scaling behavior of the start-up vortex ring}
As reported in the previous section, the circulation, the radius, and the vertical distance to the disk of the start-up vortex all reach a maximum during the motion of the disk. Besides, as shown in figures \ref{fig:III_ScalingCirculation}, \ref{fig:III_ScalingDeltaz} and \ref{fig:III_ScalingRadius}, these maxima depend on the stroke length $L$, the diameter $D$ and the travel time $\tau$ of the disk according to power laws. We demonstrate in the following that these scaling laws can be rationalized using the two-dimensional theoretical results of Wedemeyer \cite{wedemeyer_ausbildung_1961}. Wedemeyer derived with conformal mapping a theoretical approach for the self-similar growth of a vortex in the wake of a semi-infinite plate perpendicular to the $z$-direction and translating at a constant velocity $U$ in the $z$-direction from an initial starting time $t = 0$. The theoretical law for the time evolution of growth of the circulation of the vortex and its position relative to a plate of length $2H$ in the direction perpendicular to the $z$-axis is given by \cite{wedemeyer_ausbildung_1961}
\begin{equation}
    \Gamma (t) = -c_1 U H \bigg(\frac{U t}{H}\bigg)^{1/3},\quad \Delta z (t) = c_2 H \bigg(\frac{U t}{H}\bigg)^{2/3}, \mbox{~~~and~~~} \Delta r (t) = c_3 H \bigg(\frac{U t}{H}\bigg)^{2/3},
    \label{eq:GammaDxDrWedemeyer}
\end{equation}

\noindent where $c_1 \simeq 4$, $c_2 \simeq 0.4$ and $c_3 \simeq 0.11$ \cite{wedemeyer_ausbildung_1961}. In order to adapt these scalings for our present configuration, we substitute the constant velocity $U$ by the time-averaged velocity from 0 to $t^* = t/\tau$, $U(t^*) = L [1 - \cos(\pi t^*)]/2\tau t^*$, $Ut$ by the stroke length $L(t^*) = L [1 - \cos(\pi t^*)]/2$ and $H$ by the radius of the disk $D/2$. After substitution in the equation (\ref{eq:GammaDxDrWedemeyer}) for $\Gamma(t)$, we obtain the law governing the growth of the circulation
\begin{equation}
    \Gamma (t^*) =  c_\Gamma(t^*) L^{4/3}D^{2/3} /\tau,
    \label{eq:GammaAdapted}
\end{equation}

\noindent where $c_\Gamma(t^*)$ = $c_1[1-\cos(\pi t^*)]^{4/3}/4t^*$.

We make the same substitutions for $\Delta z(t)$ and $\Delta r(t)$ in equations (\ref{eq:GammaDxDrWedemeyer}) to find the theoretical prediction for the vertical and radial distances $\Delta z$ and $\Delta r$ between the centroid of the vortex and the edge of the disk:
\begin{equation}
    \Delta z (t^*) = c_z(t^*) L^{2/3}D^{1/3} ,\ \ \ \mathrm{and}\ \ \ 
    \Delta r (t^*) = c_r(t^*) L^{2/3}D^{1/3} ,
    \label{eq:dydxAdapted}
\end{equation}
where $c_z$ = $c_2[1-\cos(\pi t^*)]^{2/3}/2$ and $c_r$ = $c_3[1-\cos(\pi t^*)]^{2/3}/2$. In addition to equations (\ref{eq:GammaAdapted})-(\ref{eq:dydxAdapted}) derived from the theoretical approach of Wedemeyer, our experimental and numerical data suggest that the maximum radius of the vortex is proportional to $\Delta z_m$, $ a_m = 0.48~\Delta z_m$ so that 
\begin{equation}
    a(t^*)= c_a(t^*) L^{2/3} D^{1/3},
    \label{eq:aAdapted}
\end{equation}

\noindent where $c_a(t^*) \simeq 0.48~c_z(t^*)$. 

These theoretical results can be used for finite-size body as long as the size of the vortex is small compared to the size of the body, which is the case in the present study as $a_m/D \in [0.015,~0.15]$. In addition, a comparison with the theoretical study \cite{wedemeyer_ausbildung_1961} will only be relevant in the generation process, \textit{i.e.}, as long as the vortex grows.

\begin{figure}[t]
    \centering
    \includegraphics[width=\linewidth]{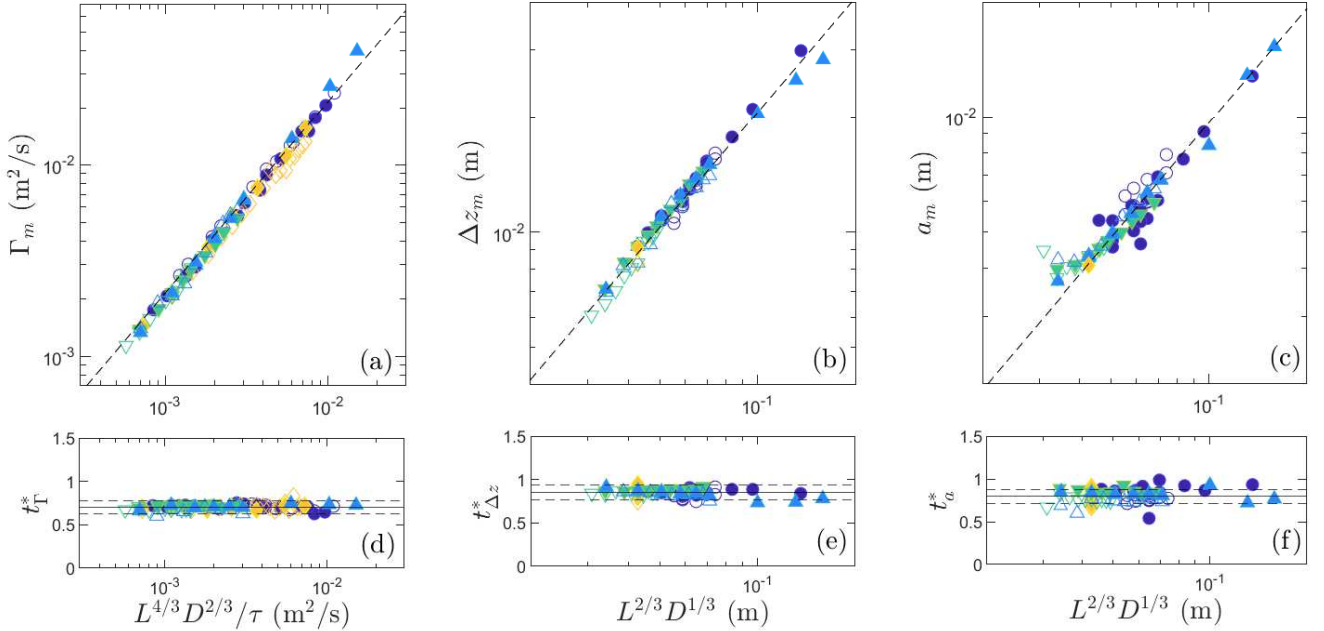}
    \caption{(a) Maximum circulation of the starting vortex $\Gamma_m$ as a function of $L^{4/3} D^{2/3} \tau^{-1}$. (b) Maximum vertical position $\Delta z_m$ and (c) the maximum core radius $a_m$ as a function of $L^{2/3} D^{1/3}$. (d-f) Times at which (d) the maximum circulation $t^*_\Gamma$ is reached, (e) the maximum vertical position $t^*_{\Delta z}$ is reached, and (f)  the maximum radius $t^*_a$ is reached when varying the stroke length $L$, the diameter $D$, and the travel time $\tau$ of the disk. The lines in (a)-(c) correspond to the power laws: (a) $\Gamma_m = c_\Gamma L^{4/3} D^{2/3}/\tau$ with $c_\Gamma \simeq 2.1$, (b) $\Delta z_m = c_z L^{2/3} D^{1/3}$ with $c_z \simeq 0.2$, and (c) $a_m = c_a L^{2/3} D^{1/3}$ with $c_a \simeq 0.1$. For the radius and the vertical position, when $\tau$ varies, only one point was represented in figures (b)-(c), which corresponds to the averaged radius (or averaged vertical position) as it does not vary with $\tau$. The continuous lines in (d)-(f) correspond to the mean value of the time of maximum and the dotted lines to the mean value $\pm$10\%. Empty and full symbols correspond to experimental and numerical data respectively.}
    \label{fig:III_AllScaling}
\end{figure}

According to equation (\ref{eq:GammaAdapted}), at a given dimensionless time, the circulation is proportional to $L^{4/3} D^{2/3}/\tau$, in agreement with the scaling laws obtained in figure \ref{fig:III_ScalingCirculation}. From equations (\ref{eq:dydxAdapted}) and (\ref{eq:aAdapted}), we also find that the vertical position of the vortex $\Delta z$ and its core radius $a$ are proportional to $L^{2/3} D^{1/3}$ and are independent of $1/\tau$ at a given non-dimensional time $t^*$, also in agreement with the observations reported in figures \ref{fig:III_ScalingDeltaz} and \ref{fig:III_ScalingRadius}. The scaling laws from equations (\ref{eq:GammaAdapted}), (\ref{eq:dydxAdapted}) and (\ref{eq:aAdapted}) are applicable for a given non-dimensional time $t^*$. Hence, they could hold for the maximum values of the circulation, radius, and vertical position as long as the time at which the maxima are reached does not vary significantly. To investigate this beyond the one-dimensional parameter range already reported in table \ref{tab:IndividualSymbols}, an additional set of experimental and numerical parameters reported in table \ref{tab:MoreDataSetsSymbols} have been made. The scaling laws derived above and the dimensionless time $t^*$ at which each maximum is reached are compared in figure \ref{fig:III_AllScaling} varying all control parameters listed in tables \ref{tab:IndividualSymbols} and \ref{tab:MoreDataSetsSymbols}.

\begin{table}[t]
\centering
\setlength{\tabcolsep}{12pt}
\begin{tabular}{ c c c c c c c} 
\multicolumn{7}{c}{}\\
\hline \hline
$L$ (cm) & $D$ (cm) & $\tau$ (s) & $L/D$ & $Re$ & Experiments & Simulations\\
\hline \hline
5.2 & 6.25-15 & 0.5-2.5 & 0.35-0.8 & $2\times 10^3$- $2 \times 10^4$ & \markerLFiveTwo & \\

2.8-10 & 7.5-30 & 0.5-2.5 & 0.09-0.7 & $2.6 \times 10^3$- $3 \times 10^4$ & & \markerLFiveTwoSimu \\

\hline \hline 
\end{tabular}
\caption{Sets of the experimental and numerical parameters and non-dimensional numbers used in this study with the corresponding data symbols used in figures.}
\label{tab:MoreDataSetsSymbols}
\end{table}

In figure \ref{fig:III_AllScaling}(a), the maximum circulation $\Gamma_m$ is reported as a function of $L^{4/3}D^{2/3}/\tau$ for all the sets of experimental and numerical parameters reported in tables \ref{tab:IndividualSymbols} and \ref{tab:MoreDataSetsSymbols}. The data collapse on a master curve of coefficient $c_\Gamma \simeq 2.1$. The maximum dimensionless time $t^*_{\Gamma}$ at which the maximum circulation is reached is shown in figure \ref{fig:III_AllScaling}(d), and does not vary significantly around the mean value $t^*_{\Gamma} =  0.70 \pm 0.02 $ for all the parameters considered here. This result highlights that the dynamics of the formation of the vortex ring are the same for the different parameters. Moreover, it indicates that Wedemeyer's scaling approach is applicable in the present axisymmetric case for the maximum circulation. The theoretical value derived from equation (\ref{eq:GammaAdapted}) gives $c_\Gamma(t^* = 0.7) \simeq 2.6$. The value obtained by fitting the experimental and numerical data ($c_\Gamma \simeq 2.1$) is slightly smaller than the theoretical one. The axisymmetry and unsteadiness of the problem can explain this discrepancy. 

The same analysis is performed for the maximum vertical position of the vortex ring $\Delta z_m$ in figure \ref{fig:III_AllScaling}(b) where $\Delta z_m$ is reported as a function of $L^{2/3} D^{1/3}$ to test equation (10). For the set of parameters for which the travel time $\tau$ varies (third row in table \ref{tab:IndividualSymbols}), only the mean value is displayed for clarity as $\Delta z_m$ is independent of $\tau$. The data collapse on a linear curve of coefficient $c_z \simeq 0.2$. The maximum dimensionless time $t^*_{\Delta z}$ at which the maximum is reached is shown in figure \ref{fig:III_AllScaling}(e) below. Similarly to the circulation, $t^*_{\Delta z}$ does not vary much around its average value $t^*_{\Delta z} = 0.85 \pm 0.04$ for all parameters considered here, making the comparison with the theoretical results applicable. From equation (\ref{eq:dydxAdapted}), we obtain that $c_z(t^* = 0.85) \simeq 0.3$, thus again slightly larger than the value we obtained $c_z \simeq 0.2$.

Finally, the maximum radius $a_m$ is plotted in figure \ref{fig:III_AllScaling}(c) as a function of $L^{2/3} D^{1/3}$. The results gather on a linear curve of coefficient $c_a \simeq 0.1$. The dimensionless time $t^*_a$ does not depend significantly on the control parameters, and its average value is $t^*_a = 0.79 \pm 0.07$. We cannot perform a comparison with a theoretical coefficient because the scaling law derived in equation (\ref{eq:aAdapted}) comes from the hypothesis that the radius is proportional to the vertical position of the vortex. There are no quantitative coefficients that come with this hypothesis. However, the assumption made works well, as the experimental and numerical results agree with the power law.

In summary, by adapting the theoretical results of Wedemeyer \cite{wedemeyer_ausbildung_1961}, we find scaling laws that capture convincingly the results. The numerical coefficients $c_\Gamma$ and $c_z$ are a little different from the ones that arise from Wedemeyer analysis which can be explained by the differences between the situation considered here and the one coming from the theoretical approach \cite{wedemeyer_ausbildung_1961}. One difference is the imposed velocity  which is not constant over time but corresponds to sinusoidal acceleration and deceleration phases in the present configuration contrary to  Wedemeyer's approach which considered a step function of the velocity. But the main difference is that the flow in our configuration is axisymmetric, and the curvature of the vortex ring generates effects that are not accounted for in the theoretical approach. The vortex ring has a self-induced velocity that affects its position relative to the disk. Moreover, the strength of the vortex sheet generated by a disk is not the same as the one generated by a 2D plate \cite{saffman_1993}. The axisymmetric configuration of a disk clearly results in significantly smaller $c_\Gamma$ and $c_z$ coefficients when compared to the plate configuration as detailed in Appendix \ref{appendix:2DSimu}.

The main discrepancy lies in the precise values of the coefficients $c_\Gamma$ and $c_z$, which are found to be smaller in the experiments and simulations. These discrepancies can be explained by the differences between the situation considered here and the one coming from the theoretical approach \cite{wedemeyer_ausbildung_1961}. Specifically, the translating velocity of the disk in the present configuration corresponds to  sinusoidal acceleration and deceleration phases, and therefore, is not constant over time, whereas Wedemeyer's approach involved a step function of the velocity. Consequently, the values of the coefficients are expected to change for a different time evolution of the disk velocity. In addition, a significant difference is that the flow in our configuration is axisymmetric, and the curvature of the vortex ring generates effects that are not accounted for in the theoretical approach. The vortex ring has a self-induced velocity that affects its position relative to the disk. Moreover, the strength of the vortex sheet generated by a disk is not the same as the one generated by a 2D plate \cite{saffman_1993}. All these geometrical differences can also explain changes in the prefactors (see more details Appendix \ref{appendix:2DSimu}).

No proper scaling laws appear from the experimental and the numerical results for the radial distance $\Delta r$ between the edge of the disk and the vortex centroid. However, equation (\ref{eq:dydxAdapted}) indicates that $\Delta r$ should vary as $L^{2/3}D^{1/3}$. The experimental and numerical results suggest (see figure \ref{fig:CompExpSimu}(c) for instance) that $\Delta r$ does not vary much with $t^*$, unlike the other features. This is believed to be a consequence of the axisymmetry of the problem. Indeed, the vortex that has formed is ring of radius $R_\text{ring}$ (see notation in figure \ref{fig:IIA_SetUpExpAndVortex}) that cannot vary much. If $\Delta r$ changes, it implies that the radius of the vortex ring $R_\text{ring}$ changes. However, by mass conservation of the fluid in the vortex ring, $a^2 R_\text{ring}$ must be constant, and so the core radius of the vortex ring should change accordingly. This relationship between the core radius and the ring radius of the vortex is believed to have a strong effect on the radial position of the vortex ring. To check this hypothesis, we perform a comparison of the scaling laws between 2D Cartesian and 2D axisymmetric simulations (see Appendix \ref{appendix:2DSimu}). The maximum radial distance $\Delta r$ of the vortex with the edge of the plate (2D Cartesian simulations) reported in figure \ref{fig:2D_Axi_Comp}(d) follows the prediction of equation (\ref{eq:dydxAdapted}) whether it is not the case for the 2D axisymmetric simulation showing that the axisymmetry of the problem indeed plays a key role on the radial position of the vortex ring.

Thanks to the scaling laws validated from a large set of experiments and numerical simulations, it is possible to predict the maximum circulation, vertical distance and core radius of the vortex from the system parameters $L$, $D$, and $\tau$. In the following, we will provide some results explaining the behavior of the vortex ring after its generation phase.

\subsection{Second phase: Reduction of circulation and radius}
\begin{figure}[t]
    \centering
    \includegraphics[width=\linewidth]{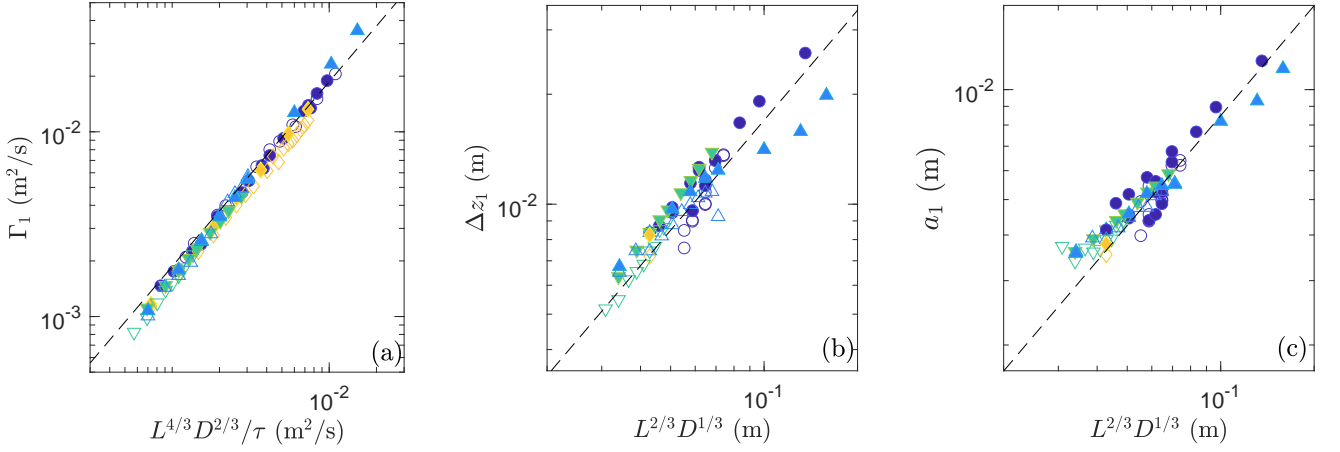}
    \caption{(a) Circulation $\Gamma_1$, (b) vertical position $\Delta z_1$ and (c) core radius $a_1$ of the start-up vortex at $t^* = 1$ when varying the stroke length $L$, the diameter $D$, and the travel time $\tau$ of the disk. For the radius and the vertical position, when $\tau$ varies only one point is represented, which corresponds to the averaged radius (or averaged vertical position) as it does not vary with $\tau$. The solid lines correspond to the power laws: (a) $\Gamma_1$ = $c_{\Gamma,1} L^{4/3}D^{2/3}/\tau$ with $c_{\Gamma,1} \simeq 1.9$, (b) $\Delta z_1$ = $c_{z,1} L^{2/3}D^{1/3}$ with $c_{z,1} \simeq 0.17$, and (c) $a_1$ = $c_{a,1} L^{2/3} D^{1/3}$ with $c_{a,1} \simeq 0.085$.}
    \label{fig:III_AllScalingtTau1}
\end{figure}

In figure \ref{fig:CompExpSimu}, we observe that the circulation $\Gamma$ and the core radius $a$ of the vortex ring start to decrease while the disk is decelerating. This decrease in circulation is due to the creation of opposite vorticity in the boundary layer between the vortex and the disk. This opposite vorticity, produced by the roll-up velocity of the vortex, penetrates the vortex, reducing its global circulation. This phenomenon has already been observed in experiments devoted to the formation of vortex rings at the outlet of a tube by a piston \cite{maxworthy_experimental_1977,didden_formation_1979} where the phase of formation is followed by a decrease of the circulation before the detachment of the vortex from the walls. We report the scaling laws obtained for the circulation $\Gamma_1$, vertical position $\Delta z_1$, and core radius $a_1$ of the starting vortex ring at $t^*=1$ when the disk stops, in figures \ref{fig:III_AllScalingtTau1}(a)-(c). We observe that scaling laws similar to the ones in figure \ref{fig:III_AllScaling} capture well the feature of the vortex ring, with a change of prefactors. Indeed, we find $c_{\Gamma,1} \simeq 1.9$, meaning that the circulation has reduced by approximately 11\% from its maximal value. The coefficient for the vertical position is $c_{z,1} \simeq 0.17$ which means that the vortex approaches the disk. Finally, $c_{a,1} \simeq 0.085$, so that the core radius of the vortex has decreased by 11\%. Note that, during the second phase, the radial position of the vortex ring changes as the vortex starts to move around the disk, hence the major radius of the ring also expands.

\subsection{Third phase: Evolution after stopping of the disk}
\begin{figure}[t]
    \centering
    \includegraphics[width=0.7\linewidth]{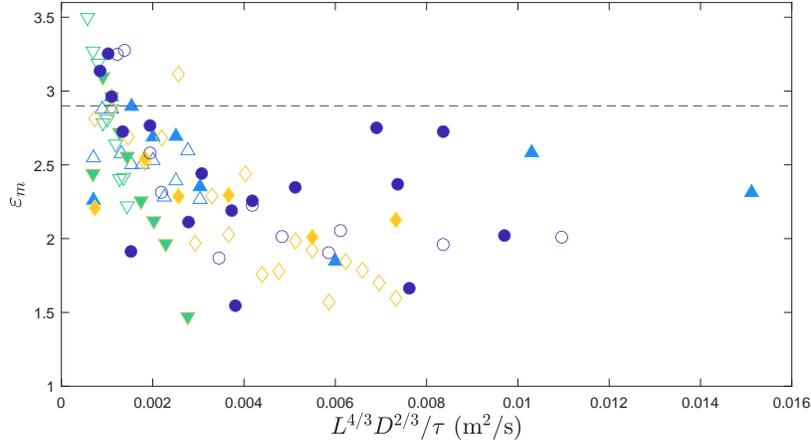}
    \caption{Maximum ellipticity $\varepsilon_m$ of the start-up vortex as a function of its maximum circulation $L^{4/3}D^{2/3}/\tau$. The solid line corresponds to the critical value $\varepsilon_c = 2.9$ \cite{moore_saffman_1971}.}
    \label{fig:EpsilonMax}
\end{figure}

Finally, after the disk has stopped ($t^* > $ 1), a counter-rotating stopping vortex ring forms at the edge of the disk, due to the roll-up velocity of the start-up vortex (see figure \ref{fig:III_VorticitySnapshots} from $t^* = 1$). This secondary vortex ring creates a strain field that can deform the initial vortex, causing it to become elliptical and affecting its dynamics. 
The ellipticity $\varepsilon$ of the initial start-up vortex defined in equation (\ref{eq:ellipticity}) is observed to increase after the disk has stopped, to reach a maximum value $\varepsilon_m$ and thereafter to decrease. 
In figure \ref{fig:EpsilonMax}, we report the maximum ellipticity $\varepsilon_m$ of the start-up vortex as a function of $L^{4/3}D^{2/3}/\tau$, which is proportional to its maximum circulation (see figure \ref{fig:III_AllScaling}(a)). 

The inviscid solution for a steady 2D vortex patch of uniform vorticity subjected to an in-plane 2D strain field derived by Moore and Saffman \cite{moore_saffman_1971} and extended to quasi-steady viscous vortices~\cite{delbende_rossi_2009} indicates that the ellipticity is an increasing function of the ratio between external strain and internal vorticity. 
Such results may also hold for vortex rings when the core size is small with respect to the curvature radius, which is assumed here. In figure~\ref{fig:EpsilonMax}, the largest values of the ellipticity are obtained for the smallest values of $L^{4/3}D^{2/3}/\tau$: they actually correspond to the cases of weaker vorticity, hence of large strain-to-vorticity ratio. Above a critical value $\varepsilon_c = 2.9$, a 2D vortex patch is inviscidly eroded and may even be destroyed. We observe that, for most cases, this critical value reported as a dashed line in figure \ref{fig:EpsilonMax} is not exceeded, except for the smallest values of $L^{4/3}D^{2/3}/\tau$.
When this occurs, it is only transient: the vortex ring is not destroyed and eventually becomes almost circular.

\begin{figure}[b]
    \centering
    \includegraphics[width=\linewidth]{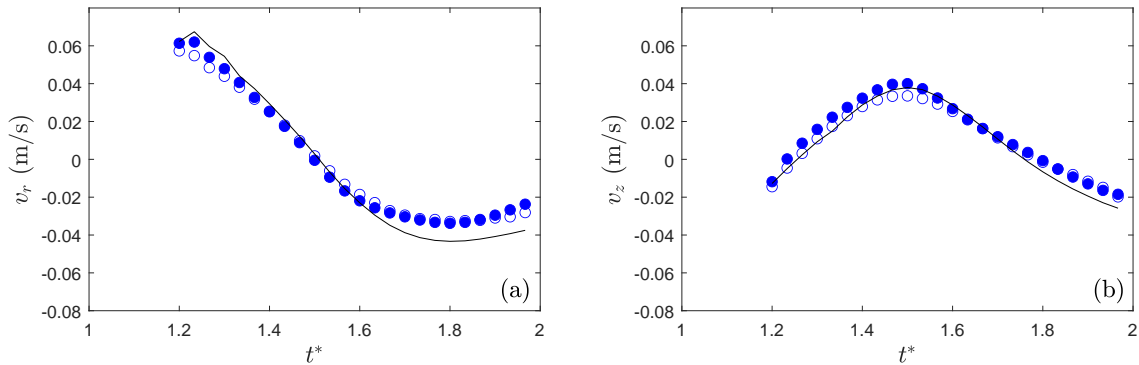}
    \caption{Time evolution of (a) the radial and (b) the vertical velocity of the start-up vortex for $L$ = 5.2 cm, $D$ = 10 cm, and $\tau$ = 1 s. Experiment (\protect\markerExpEvolution), simulation (\protect\markerSimuEvolution), and the predicted velocity (---) given by equations (\ref{eq:VRing}-\ref{eq:AllVelocity}).}
    \label{fig:III_Velocity}
\end{figure}

In addition to its deformation, the start-up vortex is subjected to its self-induced velocity and the velocity induced by the newly formed stopping vortex. The total velocity of the initial vortex ring should thus be given by the addition of its self-induced velocity in the vertical direction, which is given by \cite{saffman_1993}
\begin{equation}
    v_\text{ring} = \frac{\Gamma}{4\pi R_\text{ring}}\Bigg[\ln \bigg( \frac{8R_\text{ring}}{a} \bigg) -0.558\Bigg],
    \label{eq:VRing}
\end{equation}

\noindent and the velocity induced by the stopping vortex on the main vortex $(\dot{r}, \dot{z})$ \cite{saffman_1993}
\begin{equation}
    \begin{pmatrix}
  \dot{r}\\
  \dot{z}
\end{pmatrix}
 = \frac{\Gamma_s}{2\pi d^2}
 \begin{pmatrix}
  -(z_s-z_\mathrm{G})\\
    r_s-r_\mathrm{G}
\end{pmatrix},
\label{eq:VInduced}
\end{equation}
\noindent where $\Gamma_s$ and $(r_s, z_s)$ are the circulation and the position of the stopping vortex, $(r_\mathrm{G},z_\mathrm{G})$ is the position of the start-up vortex, and $d$ the distance between the centroids of the two vortices. An example of the temporal evolution of the features of the stopping vortex ring is given in Appendix \ref{appendix:StoppingVortex}.  The generation of the stopping vortex is found to be mainly governed by the starting vortex. Indeed, the circulation and the core radius of the stopping vortex ring follow the same scaling laws as the starting one, with prefactors decreased by almost a factor 2, as shown in Appendix \ref{appendix:StoppingVortex}.

The final velocity of the starting vortex ($v_r$, $v_z$) should thus be given by
\begin{equation}
    v_r = \dot{r},\ \ \ v_z = v_\text{ring} + \dot{z}.
    \label{eq:AllVelocity}
\end{equation}

In figures \ref{fig:III_Velocity}(a) and \ref{fig:III_Velocity}(b), the experimental (\protect\markerExpEvolution) and numerical (\protect\markerSimuEvolution) velocity components $(\dot{r_\mathrm{G}}, \dot{z_\mathrm{G}})$ of the barycenter of the main vortex rings are plotted. The expected velocity has also been plotted by substituting the different parameters in equation (\ref{eq:AllVelocity}) by their numerical value. The velocities extracted from the experimental and numerical data are in good quantitative agreement with each other and agree well with the predicted velocity given by equations (\ref{eq:VRing}-\ref{eq:AllVelocity}). Therefore, the time evolution of the position of the start-up vortex ring after the disk stops is well captured by two contributions: its self-induced velocity and the velocity induced by the stopping vortex.

\section{Conclusion}
In this study, we have investigated the properties of a vortex ring generated by the unsteady translation of a disk of diameter $D$ on a finite stroke length $L$ for Reynolds numbers ranging from $10^3$ to $2.6 \times 10^4$ and for $L/D$ ranging from 0.07 to 2. The study focused on experimental results obtained by PIV measurements and axisymmetric numerical simulations that are in good quantitative agreement. This suggests that, in the range of parameters considered here, non-axisymmetric fluctuations in the flow are not dominant in the generation of a vortex ring by the translation of a disk.

The temporal evolution of the start-up vortex ring can be described in three phases. The first stage, during which the disk accelerates, corresponds to the generation of the vortex ring. The core radius and the circulation of the vortex ring increase in time, and the vortex ring centroid moves away vertically from the disk but not radially. The maximum circulation $\Gamma_m$, core radius $a_m$, and distance $\Delta z_m$ from the disk follow scaling laws that can respectively be summarized as
\begin{equation}
    \Gamma_m \propto L^{4/3}D^{2/3}/\tau,\ \ a_m \propto L^{2/3}D^{1/3}, \mbox{~~~and~~~}  \Delta z_m\propto L^{2/3} D^{1/3}.
\end{equation} 
These scalings laws can be rationalized based on a two-dimensional theoretical approach for a semi-infinite plate animated by a constant velocity \cite{wedemeyer_ausbildung_1961}. The present study shows that the scaling laws can be applied to a vortex ring generated by a disk animated by a non-uniform velocity with a change in the prefactors.

In the second phase, the disk is still translating but decelerates so that the strength and size of the vortex ring decrease. This is due to the entrance of opposite vorticity inside the vortex ring. The loss of circulation of the vortex is estimated by 11\% in our configuration. During this phase, the vortex ring also starts to approach the disk and moves in the outward radial direction to avoid the disk.

Finally, in the last stage, after the disk has stopped, a counter-rotating stopping vortex forms at the edge of the disk due to the roll-up velocity of the primary vortex. Due to the strain field induced by the secondary vortex,the core of the main vortex deforms. In addition, the two vortices rotate in the bulk due to their mutual interaction. The displacement of the primary vortex ring is well explained by the combination of its self-induced velocity and the velocity induced by the stopping vortex.

This study focused on the behavior of the vortex ring generated in the near wake of a circular disk in unsteady translation in an unbounded fluid. In this configuration, no influence of surrounding walls has been considered. An interesting follow-up study could focus on the features of a vortex formed in the wake of a disk moving in the direction or away from a solid boundary. In addition, one can wonder what will happen if the disk is now set to oscillate continuously.

\section*{Acknowledgement}
The authors thank Johannes Amarni, Alban Aubertin, Lionel Auffray, and Rafaël Pidoux for their work on the experimental setup and to Ramiro Godoy-Diana, Fr\'ed\'eric Moisy, and Maurice Rossi for fruitful discussions.

\bibliography{bibliography}

\appendix
\section{Two-dimensional Cartesian numerical simulations}
\label{appendix:2DSimu}

\begin{table}[h!]
\centering
\setlength{\tabcolsep}{12pt}
\begin{tabular}{ |c c c c c| } 
 \hline
 $L$ (cm) & $D$ (cm) & $\tau$ (s) & $L/D$ & $Re$ \\
 \hline
2-6 & 10 & 1.67 & 0.2-0.6 & $1.9\cdot10^3$- $5.7\cdot10^3$  \\
2.8 & 5-15 & 1.67 & 0.19-0.56 & $1.3 \cdot 10^3$- $3.9\cdot10^3$  \\
2.8 & 10 & 0.25-2.5 & 0.28 & $1.8\cdot10^3$- $1.8\cdot10^4$ \\
 \hline
\end{tabular}
 \caption{Sets of simulated parameters for the 2D Cartesian numerical simulations.}
 \label{tab:2DSimu}
\end{table}

To check the effect of the axisymmetry of the disk configuration, we also performed 2D numerical simulations in a Cartesian coordinate system using the Basilisk flow solver \cite{Basilisk}. The 2D Cartesian computational domain is similar to the 2D axisymmetric one given in figure \ref{fig:SetUpNum}(a). The domain is again a square defined by $(x,z)\in [0,\lambda]\times[-\lambda/2,\lambda/2]$, where $\lambda=4D$.
The solid plate is taken into account similarly to what is done in the axisymmetric simulations, and the velocity of the plate is imposed according to equation (\ref{eq:motion_law}). At outer boundaries, no-slip conditions are used. The same parameters are used in terms of the maximum refinement level and Courant number. The adaptative refinement is again implemented to improve the spatial discretization near boundaries and high velocity gradient zones. The numerical simulations that are performed are summarized in table \ref{tab:2DSimu}.

\begin{figure}[t]
    \centering
    \includegraphics[width=0.9\linewidth]{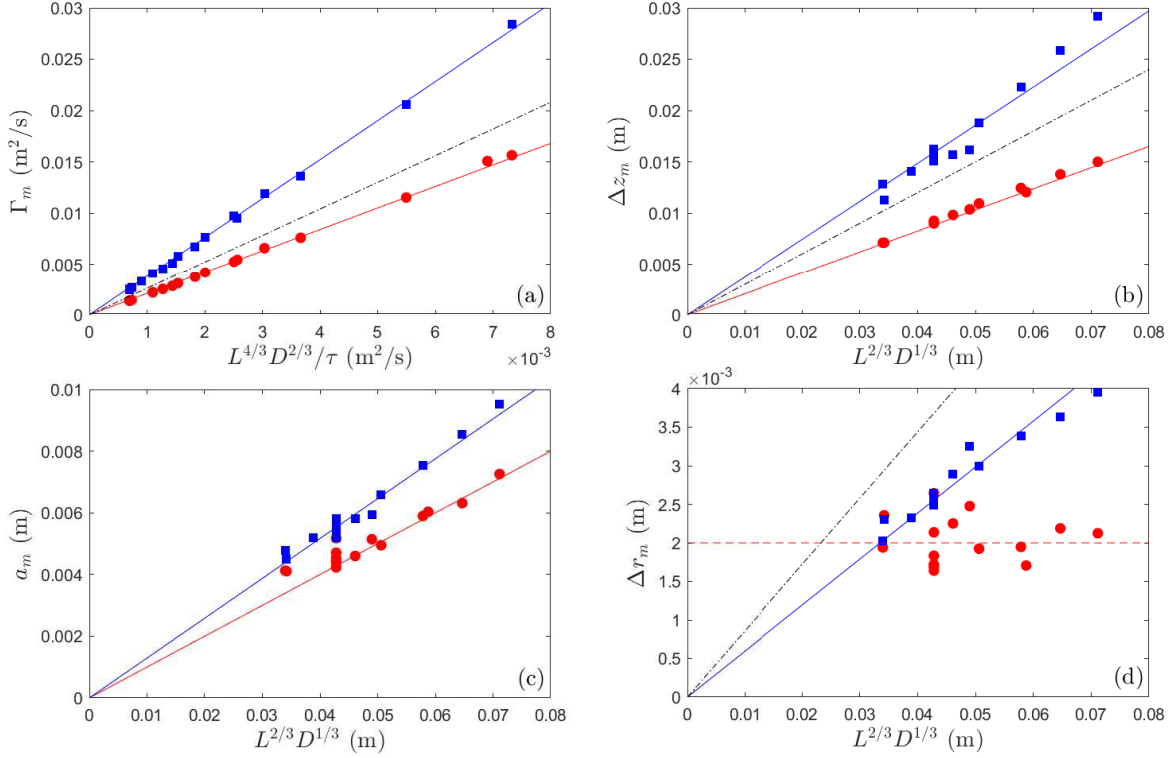}
    \caption{Comparison of 2D Cartesian (\protect\markerSimuTwoD, plate) and 2D axisymmetric (\protect\markerSimuAxi, disk) simulations showing the evolution of (a) the circulation $\Gamma_m$, (b) the vertical position $\Delta z_m$, (c) the core radius $a_m$ and (d) the radial position $\Delta r_m$ of the vortex when varying the stroke length $L$, the diameter $D$ (or the length of the 2D plate) and the travel time $\tau$ in the same figure. The solid lines correspond to the equations (9)-(11) with the following fitted coefficients for the 2D Cartesian (plate) simulations: (a) $c_{\Gamma,2D} \simeq 3.8$, (b) $c_{z,2D} \simeq 0.37$, (c) $c_{a,2D}  \simeq 0.13$ and (d) $c_{r,2D} \simeq 0.06$ and for the 2D axisymmetric (disk) simulations: (a) $c_\Gamma \simeq$ 2.1, (b) $c_z \simeq$ 0.2, and (c) $c_a \simeq $ 0.1. The red dashed line in (d) corresponds to the constant $\Delta r_m = 2.2~$mm. The black dash-dotted lines correspond to the theoretical coefficients of Wedemeyer \cite{wedemeyer_ausbildung_1961}: (a) $c_\Gamma = 2.6$, (b) $c_z = 0.3$ and (d) $c_r = 0.086$.}
    \label{fig:2D_Axi_Comp}
\end{figure}

The evolution of the characteristics of the 2D Cartesian vortex with the control parameters is summarized in figures \ref{fig:2D_Axi_Comp}(a)-(d), along with the characteristics of the vortex ring (axisymmetric case). The scaling laws developed in equations (\ref{eq:GammaAdapted})-(\ref{eq:aAdapted}) agree well with the 2D Cartesian simulations for the different parameters. In particular, $\Delta r_m$ follows the scaling laws provided in equation (\ref{eq:dydxAdapted}) unlike the axisymmetric simulations or the experimental case as seen in figure \ref{fig:2D_Axi_Comp}(d).

In addition, it can be seen in figure \ref{fig:2D_Axi_Comp}(a) that the circulation of the vortex in the 2D Cartesian simulations is larger than the circulation  of the vortex ring in the axisymmetric simulations. For the same velocity and size of the plate/disk, the strength of the vorticity sheet is larger in the 2D Cartesian case than in the 2D axisymmetric case \cite{saffman_1993}.

Moreover, the coefficient of the scaling law for the maximum radius is larger than in the axisymmetric case. Hence, the core size of the vortex in the Cartesian simulations is larger than in the axisymmetric simulations (see figure \ref{fig:2D_Axi_Comp}(c)).  Finally, the 2D vortex is going further from the disk than the axisymmetric vortex ring. In the axisymmetric case, the self-induced velocity of the vortex ring can be responsible for this difference.

In conclusion, the value of the scaling coefficients is larger in the 2D numerical simulations. In addition to this, the scaling laws derived by Wedemeyer \cite{wedemeyer_ausbildung_1961} fail to predict the radial position of the vortex ring in the 2D axisymmetric case whether they agree with the 2D Cartesian numerical simulations. 

\section{Features of the stopping vortex ring}
\label{appendix:StoppingVortex}

The knowledge of the generation and the temporal evolution of the stopping vortex ring is crucial to better understand the behavior of the starting vortex ring after the disk has stopped. For this purpose, the temporal evolution of the circulation and the core radius of the stopping vortex have been extracted from numerical simulations.

\begin{figure}[t]
    \centering
    \includegraphics[width=0.75\linewidth]{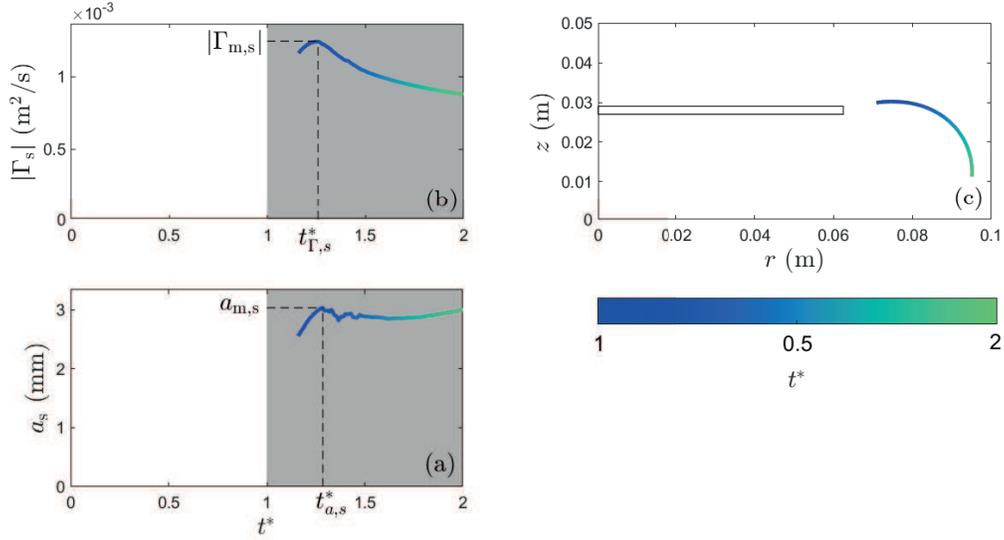}
    \caption{Time evolution of (a) the circulation $|\Gamma_s|$, (b) the radius $a_s$ and (c) the position of the barycenter of the stopping vortex from a numerical simulation for $L$ = 2.8 cm, $D = 12.5~$cm, and $\tau$ = 1.67 s.}
    \label{fig:VortexFeatures_Secondary}
\end{figure}

\begin{figure}[h]
    \centering
    \includegraphics[width=0.6\linewidth]{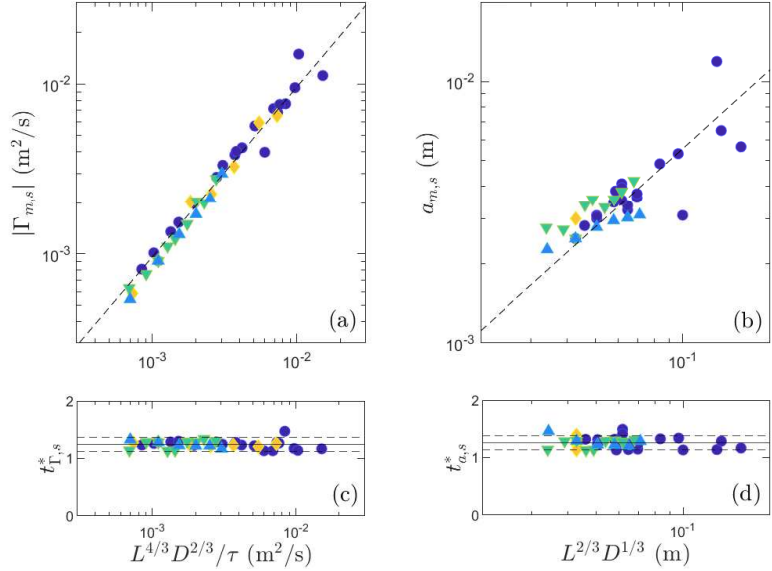}
    \caption{(a) Maximum circulation of the stopping vortex $|\Gamma_\mathrm{m,s}|$ as a function of $L^{4/3} D^{2/3} \tau^{-1}$. (b) Maximum core radius of the stopping vortex $a_\mathrm{m,s}$ as a function of $L^{2/3} D^{1/3}$. (c) Time $t^*_{\Gamma,s}$ at which the maximum circulation is reached, and (d) the time $t^*_{a,s}$ at which the maximum radius is reached. The solid lines in (a)-(b) correspond to the following power laws: (a) $|\Gamma_\mathrm{m,s}| = c_{\Gamma,s} L^{4/3} D^{2/3}/\tau$ with $c_{\Gamma,s} \simeq 1.1$, (b) $a_\mathrm{m,s} = c_{a,s} L^{2/3} D^{1/3}$ with $c_{a,s} \simeq 0.06$. The solid lines in (c)-(d) correspond respectively to the mean value of the time of maximum circulation  and of maximum radius (c) $t^*_{\Gamma,s} = 1.23$ (d) $t^*_{a,s} = 1.27$ and the dotted lines to the mean value $\pm$10\%.}
    \label{fig:GlobalScalingWithTime_Secondary}
\end{figure}

An example of the time evolution of the stopping vortex is shown in figures \ref{fig:VortexFeatures_Secondary}(a)-(c) for the same configuration as in figures 3 and 4. Its circulation and core radius are reported for $t^* > 1.16$ so that the stopping vortex is well defined and therefore easy to follow with the Matlab routine.

In figure \ref{fig:VortexFeatures_Secondary}(a), the circulation of the stopping vortex ring $|\Gamma_s|$ increases until it reaches its maximum value $|\Gamma_\mathrm{m,s}|~=~0.0012~$m$^2$/s at $t^*_{\Gamma ,s}\simeq 1.26$, followed by a gradual decreases. In figure \ref{fig:VortexFeatures_Secondary}(b), the core radius of the stopping vortex $a_s$ is found to increase with time and reaches its maximum value $a_\mathrm{m,s} = 3~$mm at $t^*_{a,s} \simeq 1.29$. Finally, figure \ref{fig:VortexFeatures_Secondary}(c) gives the position of the stopping vortex in the laboratory frame of reference. As expected, the vortex rotates in the fluid because it follows the position of the start-up vortex ring.

As for the starting vortex, the maximum circulation and core radius of the stopping vortex ring have been systematically computed and the Wedemeyer scaling laws have been tested in the figures \ref{fig:GlobalScalingWithTime_Secondary}(a)-(b). The time at which the maxima are reached $t^*_{\Gamma,s}$ and $t^*_{a,s}$ are also reported in figures \ref{fig:GlobalScalingWithTime_Secondary}(c) and (d), respectively. The parameter range was restricted to $L < 10~$cm because for longer stroke length, several small secondary vortices are shed from the disk and there is no main stopping vortex. The values of the maximum circulation of the stopping vortex plotted as a function of $L^{4/3} D^{2/3}/\tau$ in figure 14(a) collapse on a master curve of coefficient $c_{\Gamma,s} \simeq 1.1$. The maximum circulation of the stopping vortex is found to be mainly driven by the circulation of the starting vortex and it is observed to be about half the maximal circulation exhibited by the primary vortex. Moreover, the time at which the maximum circulation is reached does not vary much around the mean value $t^*_{\Gamma,s} = 1.23 \pm 0.05$. The generation of the stopping vortex ring is hence very similar in the range of parameters studied here.

The maximum core radius $a_{s,m}$ of the stopping vortex ring is plotted as a function of $L^{2/3} D^{1/3}$ in figure \ref{fig:GlobalScalingWithTime_Secondary}(b). Although the data are more scattered than for the circulation, a linear curve of coefficient $c_{a,s} \simeq 0.06$ gives a good approximation of the maximum core radius. The radius of the stopping vortex is 1.6 times smaller than the radius of the starting vortex. Moreover, the time at which its maximum core radius is reached is displayed in figure \ref{fig:GlobalScalingWithTime_Secondary}(d). This time does not vary much around its mean value $t^*_{a,s} = 1.27 \pm 0.09$, which again shows that the generation of the secondary vortex is very similar across the parameters sets.

In conclusion, the generation of the stopping vortex is mainly governed by the starting vortex. The circulation and the core radius follow the same scaling laws with only a change in numerical prefactor. The stopping vortex is found to have about the half size and the half circulation compared with the starting vortex, meaning that these two vortex rings have almost the same maximum azimuthal velocities $v_\theta$.

\end{document}